\documentstyle[epsf,psfig,12pt]{article}
\newcommand{\beq}{\begin{equation}}
\newcommand{\eeq}{\end{equation}}
\newcommand{\bea}{\begin{eqnarray}}
\newcommand{\eea}{\end{eqnarray}}
\newcommand{\ba}{\begin{array}}
\newcommand{\ea}{\end{array}}

\newcommand{\gap}{{\stackrel{_{\scriptstyle >}}{_{\scriptstyle\sim}}}}

\newcommand{\ycal}{{Y_{\rm CAL}}}
\newcommand{\xcal}{{X_{\rm CAL}}}

\newcommand{\CPC}[3]{Comput. Phys.\ Commun.\ {\bf#1} (#2) #3}

\newcommand{\PL}[3]{Phys.\ Lett.\ {\bf#1} (#2) #3}

\newcommand{\PR}[3]{Phys.\ Rev.\ {\bf#1} (#2) #3}

\newcommand{\NIM}[3]{Nucl.\ Instr.\ and Meth.\ {\bf#1} (#2) #3}

\newcommand{\PHM}[3]{Phil.\ Mag.\ {\bf#1} (#2) #3}
\begin{document}
\thispagestyle{empty}
\begin{flushleft}
hep-ex/9701015\\
DESY 97-006\\
\today
\end{flushleft}
\vspace*{\fill}
\begin{flushleft}
  \begin{LARGE}
  \begin{bf}
Design and Test of a \\
Forward Neutron Calorimeter\\
for the ZEUS Experiment\\
  \end{bf}
  \end{LARGE}
\vspace{1.5cm}
The ZEUS FNC Group\\
\vspace{0.5cm}
S.~Bhadra$^{1)}$,
I.~Bohnet$^{4)}$, 
M.~Cardy$^{1)}$,
U.~Dosselli$^{2)}$,
C.-P.~Fagerstroem$^{1)}$,
W.~Frisken$^{1)}$,
K.~Furutani$^{1)}$,
D.~Hanna$^{3)}$,
U.~Holm$^{4)}$,
K.~F.~Johnson$^{5)}$,
M.~Khakzad$^{1)}$,
G.~Levman$^{6)}$, 
J.~N.~Lim$^{3)}$, 
B.~Loehr$^{7)}$,
J.~F.~Martin$^{6)}$,
C.~Muhl$^{7)}$,
T.~Neumann$^{4)}$,
M.~Rohde$^{7)}$,
W.~B.~Schmidke$^{1)}$,
D.~G.~Stairs$^{3)}$,
H.~Tiecke$^{8)}$,
C.~Voci$^{2)}$\\
\vspace{0.5cm}
{\it
$^{1)}$ York University, Toronto, Canada$^{a}$\\
$^{2)}$ University of Padova, Padova, Italy$^{b}$\\
$^{3)}$ McGill University, Montreal, Canada$^{a,c}$\\
$^{4)}$ I. Institute of Experimental Physics, 
           University of Hamburg, Hamburg, FRG$^{d}$\\
$^{5)}$ Florida State University, Tallahassee, Florida, USA\\
$^{6)}$ University of Toronto, Toronto, Canada$^{a}$\\
$^{7)}$ DESY, Hamburg, FRG\\
$^{8)}$ NIKHEF, Amsterdam, The Netherlands$^{e}$
}
\end{flushleft}
\begin{tabular}[h]{rp{16cm}}   
$^{a}$ &  supported by the Natural Sciences and Engineering\\
       &  Research Council of Canada (NSERC)\\

$^{b}$ &  supported by the Italian National Institute for Nuclear Physics
          (INFN)\\

$^{c}$ &  supported by the FCAR of Qu\'ebec, Canada\\
$^{d}$ &  supported by the German Federal Ministry for Education and Science,\\
       &  Research and Technology (BMBF) \\

$^{e}$ &  supported by the Netherlands Foundation for Research on
          Matter (FOM)\\

\end{tabular}

\newpage
\noindent
\begin{abstract}
\noindent
A lead scintillator sandwich sampling calorimeter
has been installed in the HERA tunnel 105.6 m
from the central ZEUS detector in the proton beam
direction. It is designed
to measure the energy and scattering angle of
neutrons produced in charge exchange
$ep$ collisions. Before installation the
calorimeter was tested and calibrated in
the H6 beam at CERN where 120 GeV electrons, muons, 
pions and protons were made incident
on the calorimeter. In addition, the spectrum of
fast neutrons from charge exchange proton-lucite
collisions was measured. The design and construction
of the calorimeter is described,
and the results of the CERN test reported.
Special attention is paid to the measurement of
shower position, shower width, and the separation
of electromagnetic showers from hadronic showers.
The overall energy scale as determined from the
energy spectrum of charge exchange neutrons is
compared to that obtained from direct beam hadrons. 
\end{abstract}
\section{Introduction}

The ZEUS\cite{zeus1,zeus2} collaboration has installed a
Forward Neutron Calorimeter (FNC)
in the HERA tunnel 
105.6 m from the central detector 
in the downstream proton direction. The calorimeter,
which views the interaction point from zero degrees,
is designed to study
nucleon charge exchange reactions in $ep$ collisions
by detecting energetic neutrons produced at
small scattering angles.
The feasibility of the idea was tested successfully
with two small test calorimeters in 1993\cite{fnc1}
and 1994\cite{fnc2}.
High energy neutrons were detected in both photoproduction and
deep inelastic scattering\cite{glasgow,brkic,fnc2}.

Here we report on the
design and construction of the FNC.
We present results from the test and calibration studies
at CERN, with 
special attention paid to the measurement of
shower position, shower width, and the separation
of electromagnetic showers from hadronic showers.
At HERA the overall energy scale is determined 
from the energy spectrum of neutrons produced
by charge exchange of beam protons with residual
gas in the beam pipe. A test of this procedure was
made at CERN by comparing the neutron energy spectrum
from the charge exchange of beam protons in a lucite
target with that obtained from the direct beam.

\section {Design and Construction of the FNC}

\subsection{Introduction}

The FNC
detects neutrons 
produced at a scattering angle less than a
milliradian
and with energies as high as 
the HERA proton beam energy of 820~GeV.
In physics studies of nucleon charge 
exchange, the energy transferred from
the incoming proton
($\Delta E = E_b -E$, the difference between
the beam energy and the scattered neutron's energy)
is an important quantity.
Excellent resolution on $E$ is needed to obtain
good resolution on $\Delta E$
when $E$ is near the kinematical
end point $E_b$.
The best
energy resolution for high energy hadrons is obtained by
the use of a `compensating' calorimeter whose response to
energy deposited by electromagnetic showers equals 
its response to energy deposited by hadronic showers ($e/h = 1$).
The other physics requirement is a measurement of the
transverse momentum, $p_T$, of the neutron. This can be
achieved by transverse segmentation of the calorimeter and 
its readout.

To minimize costs, as well as development and construction time,
we chose to design the FNC as a lead-scintillator
sandwich calorimeter read out by wavelength shifting light
guides and photomultiplier tubes. 
A calorimeter of similar design
built by the ZEUS calorimeter group was shown to
compensate\cite{pbcal}. That calorimeter had a unit cell
consisting of 1.0 cm of lead and 0.26 cm of scintillator. In
hadron beams up to 100 GeV at CERN it proved
to have an energy resolution of
$\delta E/E = 0.45/\sqrt E$ with a very small constant term.
After correcting for transverse leakage of hadronic energy it had
an electron to hadron ratio $e/h$ close to unity.

The containment of 95\% of the energy of an 800 GeV hadronic 
shower requires 9-10 interaction lengths
of absorber\cite{bintinger,bock}. To ensure this
depth in the FNC
we increased the ratio of
lead to scintillator thickness from 4:1, which gives a
compensating calorimeter, to 5:1,
in order to meet the space restrictions in the HERA
tunnel (see section~\ref{description}).
As a result, we expect the FNC to overcompensate ($e/h<1$), at
least for energies below 100 GeV; however, 
Monte Carlo studies\cite{teresa} with FLUKA indicate 
that the $e/h$ ratio depends only weakly on
the sampling fraction. Furthermore, 
FLUKA predicts a weak dependence
of $e/h$ on the neutron energy. 
Hence the FNC should be close
to compensating throughout the energy range 
up to 820 GeV.

Since the sampling term
in the energy resolution function increases
as the square root of
the sampling fraction we expect 
a degradation of the energy resolution by 12\%. 
More serious
is the degradation due to transverse energy
leakage. The Monte
Carlo studies showed
that the energy resolution of the calorimeter
would be degraded to 60-70\%/$\sqrt E$. 
This large sampling term will dominate the energy
resolution even at the HERA kinematical end point
of 820 GeV provided the constant term is held to 2\% or
better. 

\subsection{Description of the Calorimeter}
\label{description}

\begin{figure}[htbp!]
\psfig{file=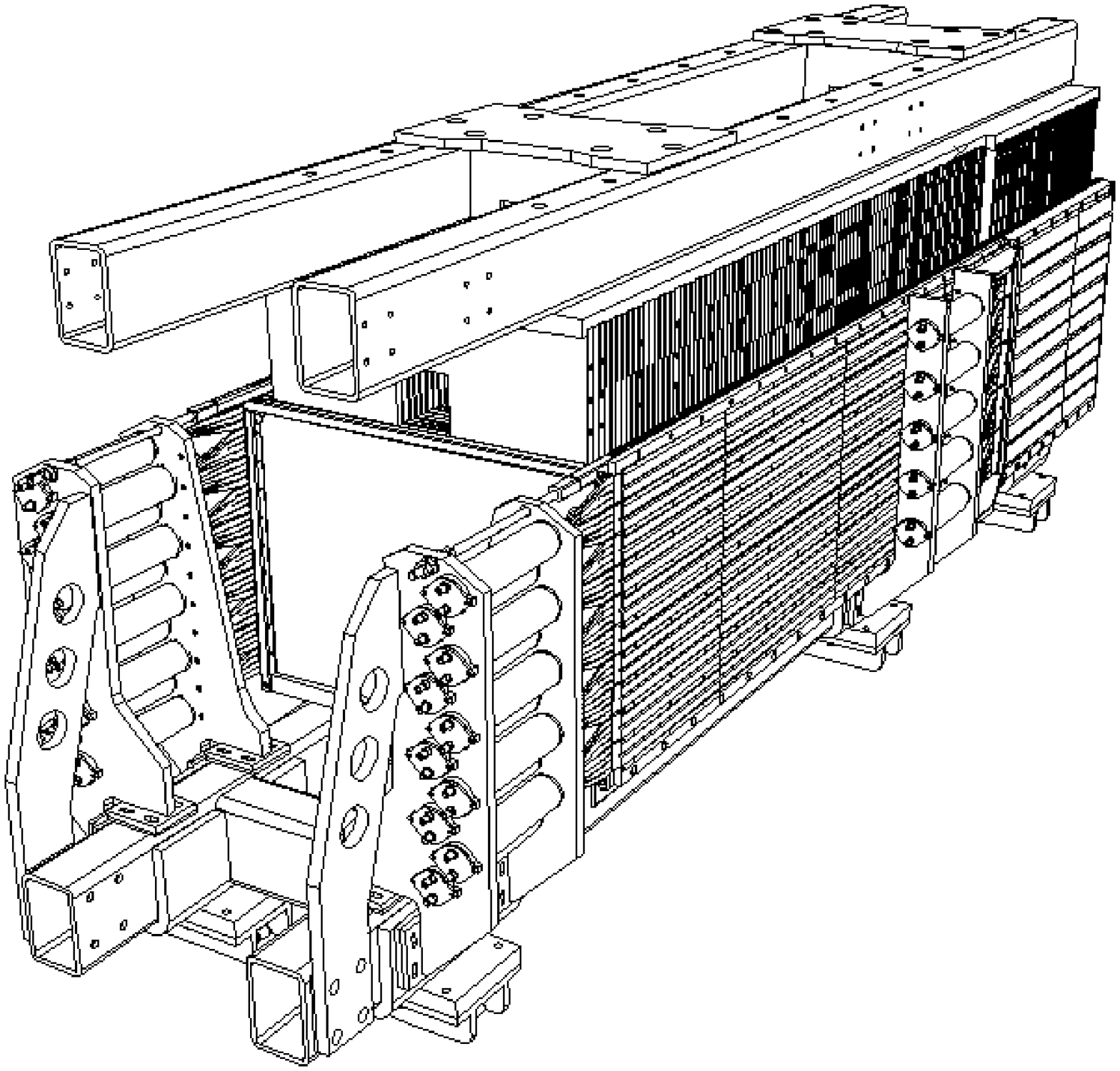,height=14cm,angle=-90}
\caption{An isometric view of the FNC showing the top and 
bottom front and rear sections with their photomultiplier
tube assemblies. At HERA the neutron line of flight, from
the bottom left to the top right, enters the calorimeter 16 cm
below the top of the bottom section.
The top section was not present for the
CERN tests.}
\label{fnc3_diagram}
\end{figure}

\begin{figure}[htbp!]
\epsfysize=14cm
\centerline{\epsffile{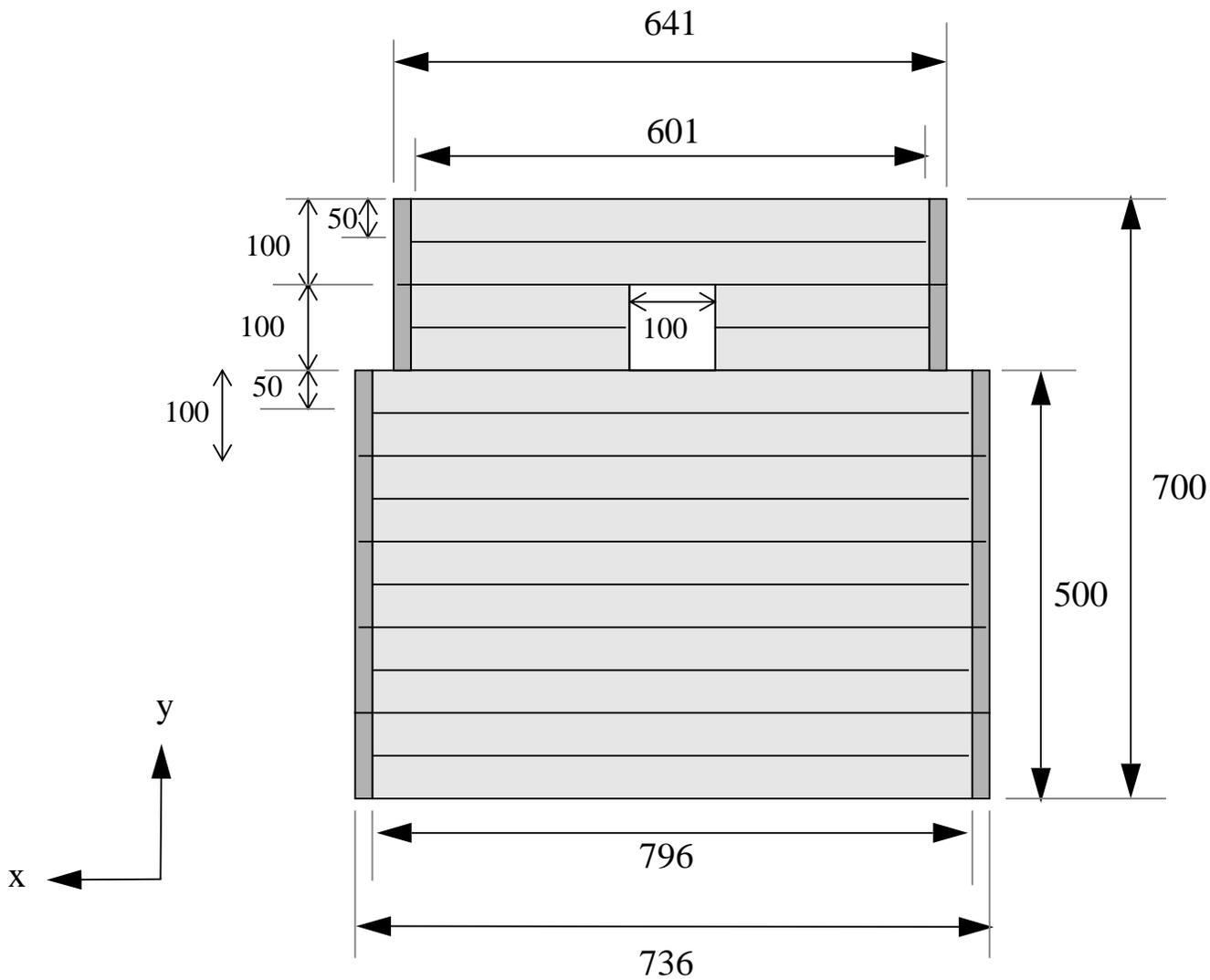}}
\caption{A schematic view of the front of the FNC.
For the CERN test,
the front towers in the bottom part were labeled
from 1 to 10, starting at the bottom. 
All dimensions are in mm.}
\label{fnc_view1}
\end{figure}

The FNC, built in two parts 
(Figs.~\ref{fnc3_diagram} and~\ref{fnc_view1}),
occupies a very restricted space in
the HERA tunnel between the electron beam line and the
tunnel wall, after the 6 mrad upward bend of the
protons so that they can circulate above the
electrons. It lies immediately in front of 
the beginning of the cold section of the proton machine.
The available space
beneath the proton beam pipe is limited in height 
to 50 cm.
The zero degree line enters the
calorimeter 16 cm below the beam pipe. 
The top part of height 20~cm
was constructed to cap the bottom part by
surrounding the beam pipe.
The two parts together provide symmetrical
up-down coverage, 35~cm in diameter.
The total length, including
readout and scintillation counters, is limited
to 3~m.
Since space is needed for PMT assemblies and scintillation
counters both before and after the calorimeter, only
about 2~m was available for the active calorimeter itself. 

The calorimeter is 134 layers deep with
a unit sampling cell as given in
Table~\ref{celltable}. 
The energy sampling fraction is 3\% for minimum ionizing
particles (mip) and 2\% for electrons,
as determined using GEANT.
For readout, it is divided longitudinally
into a front section
of 95 layers and a rear `tail catcher' of 39 layers.
This division serves two purposes. By shortening
the length of the wavelength shifter,
it reduces the difficulty of ensuring
the longitudinal uniformity of response
(section~\ref{masks}).
By having a shower tail catcher, late starting showers
which have increased probability of large
longitudinal energy leakage can be tagged.
Since the rear is not needed
for position measurements, its transverse
segmentation is reduced.
\begin{table}[htbp!]
\begin{center}
\begin{tabular}{||c|c|c||} \hline\hline
Material        & Depth      & Absorption Lengths\\ 
                & (cm)       & $\lambda_A$ \\ \hline
Pb              & 1.25       & 0.073 \\ \hline
scintillator    & 0.26       & 0.003 \\ \hline
paper           & 0.04       & 0\\ \hline 
air             & 0.1        & 0\\ \hline\hline
total           & 1.65       & 0.076 \\ \hline\hline
134 layers      & 221.1      & 10.2 \\ \hline\hline
\end{tabular}
\end{center}
\caption{The structure of a unit cell of the FNC.}
\label{celltable}
\end{table}

The calorimeter is read out on both sides by wavelength
shifting light guides (WLS) and photomultiplier tubes (PMT).
The sections of the
calorimeter, namely, 
bottom front and rear, and top front and rear,
have different widths. The
maximum width of 74 cm was constrained by the presence 
of electron
quadrupoles and radio-frequency cavities.
Table~\ref{segtable} gives the nominal size of each
section of the FNC.
\begin{table}[htbp!]
\begin{center}
\begin{tabular}{||c|c|c|c|c|c|c||} \hline\hline
Section & layers & length & depth & height x width & tower size
       & PMT channels \\ 
  & & (cm) & $\lambda_A$ & (cm x cm) & (cm) & \\ \hline\hline
Bottom front & 95 & 156.75 & 7.2 & 50 x 70   &  5 & 20 \\ \hline
Bottom rear  & 39 &  64.35 & 3.0 & 50 x 74   & 10 & 10  \\ \hline\hline
Top front    & 95 & 156.75 & 7.2 & 20 x 60.5 &  5 &  8 \\ \hline
Top rear     & 39 &  64.35 & 3.0 & 20 x 64.5 & 10 &  4 \\ \hline\hline
\end{tabular}
\end{center}
\caption{The size of the FNC sections. All the scintillator strips
have a true height of 4.94 cm.}
\label{segtable}
\end{table}

The calorimeter readout is divided into
vertical towers to permit
a measurement of the impact position of
the scattered neutron on the face of the
calorimeter. This position is needed
for the measurement of
$p_T$ of the scattered neutron.
The vertical position, $y$, is determined
by energy sharing between towers;
the horizontal position, $x$, 
by light (collected charge) division
between the left and right sides;
the HERA proton
beam has an average angular spread at the interaction
point of about 70 $\mu rad$. This results in an
intrinsic spread in the neutron impact position
of 0.7 cm, independent of its $p_T$. This intrinsic
spread sets the limit for position resolution.
With the assumption that the position
resolution scales as $1/\sqrt E$,
and the requirement that
neutrons with $E \gap$ 200 GeV have 
a position resolution better than
0.7 cm, we demand a  
resolution of 10 cm/$\sqrt E$.
Experience with 
the uranium calorimeter\cite{ucal} in ZEUS
indicates that this can be achieved,
vertically, with tower segmentation of
5 cm. The scintillator, SCSN-38, was cut into strips
4.94 cm high, and wrapped in
white Tyvek paper.
All the scintillator strips
have the same height, including the ones in the
rear section where vertical segmentation 
of the readout is 10 cm.
Here each wavelength
shifting light guide reads out 
two rows of scintillator strips.

The horizontal position resolution depends on 
$n_{pe}$, the
number of photoelectrons collected, and on $\lambda$,
the attenuation length of light in
the scintillator strips. 
There is a contribution to the position resolution from
shower fluctuations. For electrons, this is much less than
the photostatistics contribution, while for hadrons we expect
the fluctuation component to be larger, but the photostatistics
component still dominant.
The error on the horizontal position measurement is then
$\delta x = \lambda/\sqrt{n_{pe}}$\cite{ucal}. 
The ZEUS uranium calorimeter collects 220 pe/GeV, and
ZEUS tests\cite{test60}
suggest that $\lambda$ would be about 100 cm
for the 70 cm~x~5~cm~x~0.26~cm scintillator strips.
By assuming that the light collecting components
of the FNC have similar optical properties to 
those of the ZEUS calorimeter, we estimate
$n_{pe}$ for the FNC from the ZEUS calorimeter
number. Since the FNC has
a smaller sampling fraction (3:9), longer tower
widths (70:20), and transmission rather than backing
masks for the WLS, 
we estimated the FNC would collect only 45 pe/GeV,
for a resolution in $x$ of about 15 cm/$\sqrt E$.

\subsection{Wavelength Shifter and Photomultipliers}
\label{masks}

An excellent
wavelength shifting light guide material
to match the SCSN-38 scintillator is 
Y7 in PMMA base (30 ppm doping);
however, the manufacturer, 
Kuraray\footnote{Kurary Co., Tokyo, Japan.}, 
had only enough stock on hand for the rear section
of the calorimeter, so the front section light guides
were produced from similar material (BC-482A with 20 ppm doping) 
manufactured by Bicron\footnote{Bicron Corporation, Newbury Ohio, USA.}.

The WLS in FNC were made 0.4 cm thick
to absorb as much as possible
the light leaving the scintillator strips
while still having the light guides match the size of
the PMTs.
So that longitudinal shower fluctuations 
do not degrade the energy response of the calorimeter,
transmission masks were positioned between the WLS and the 
scintillator
to correct for longitudinal attenuation of light
along the WLS.
The masks 
were a pattern of black lines
printed on 3M overhead projector transparencies
using a 300 dpi laser printer. 

\begin{figure}[htbp!]
\epsfysize=14cm
\centerline{\epsffile{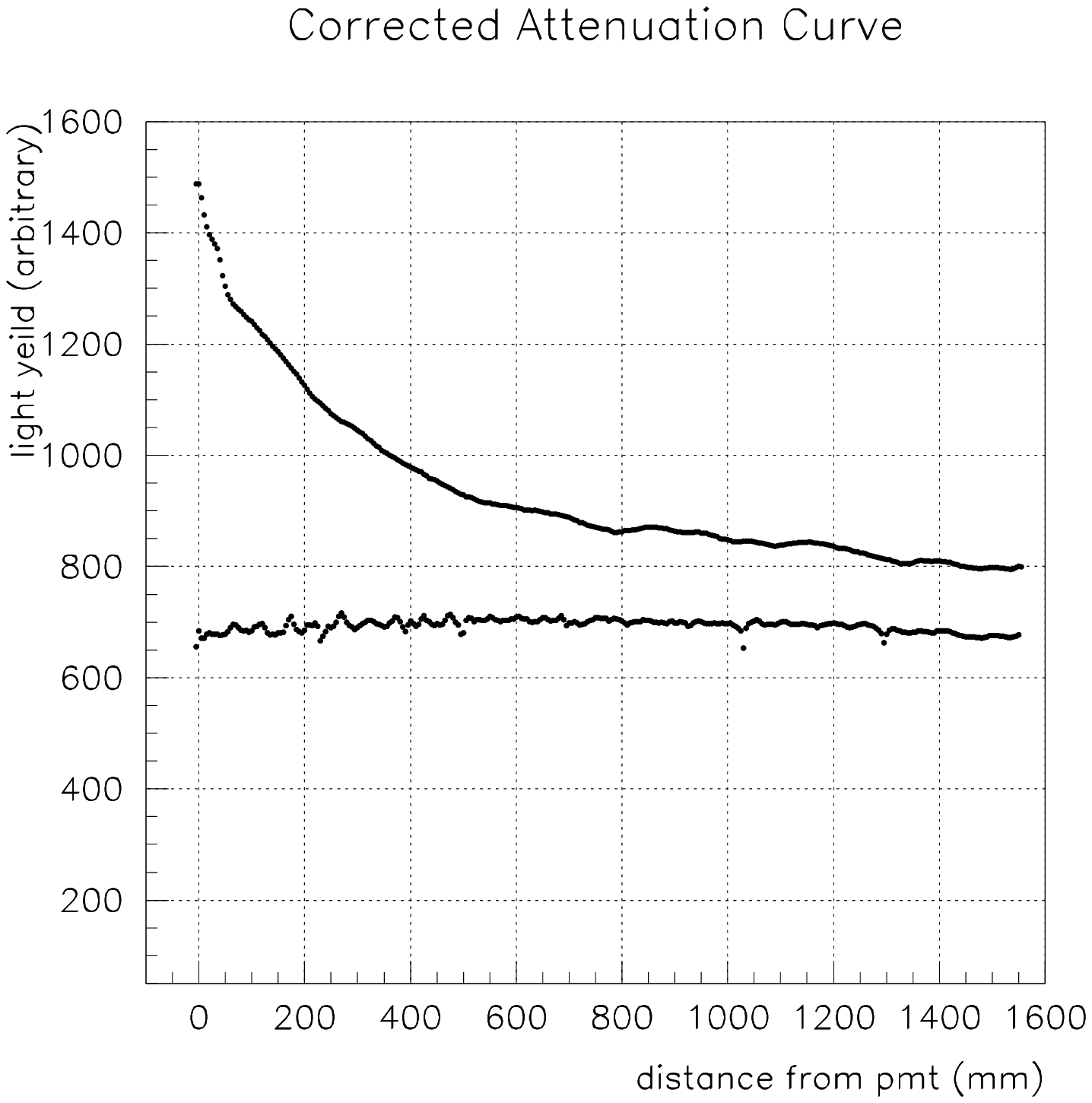}}
\caption{The top curve is the attenuation of a typical 
front wavelength shifting light guide before correction 
with a transmission mask. The
bottom curve is the same light guide after correction.}
\label{baf}
\end{figure}

The attenuation lengths of the light guides were measured
using equipment
and procedures developed for the ZEUS calorimeter\cite{zeus2}.
Figure~\ref{baf}
shows a comparison of the
curves, before and after masking, for a sample light guide. 
For the front WLS the masking reduces the peak light yield at
the PMT 
by 2.2; for the rear, the reduction is by 1.4.
The transparency material itself 
accounts for some of this reduction since,
as measured by a spectrophotometer, it
transmits only 79\% of incident light at a wavelength of 420 nm
where the emission spectrum of SCSN-38 peaks. 

After masking the response should be flat. 
A measure of the residual non-uniformity is
the root mean square (rms) 
deviation from a constant of the corrected light
curves. This averaged 2\% of the mean. 

Two types of photomultiplier tube were used:
Phillips XP 2008/UB and Valvo XP 2011/03.
Fifty pairs of tubes and resistive bases,
which were spare from the ZEUS calorimeter
tests at CERN, were obtained and tested
in the ZEUS photomultiplier
tube test facility to determine their
characteristic gain and linearity.

\subsection{$^{60}\rm\bf Co$ Source Scans}

We studied the FNC by scanning it with a 1 mCi $^{60}$Co
source, 
a procedure developed by the ZEUS calorimeter group
\cite{co60}.
This allowed us to check the
quality of the transmission masks, 
to look for construction problems,
and to measure the relative gain of each channel.

A mechanical assembly equipped with a stepping motor
moved the source
through brass guide tubes fixed outside the calorimeter
along the WLS.
An IBM 486 PC containing an integrating ADC card
controlled the stepping motors and measured
current from the two PMTs viewing the tower along which
the source was moving.
Readings were taken about 0.15 cm apart.

\begin{figure}[htbp!]
\epsfysize=14cm
\centerline{\epsffile{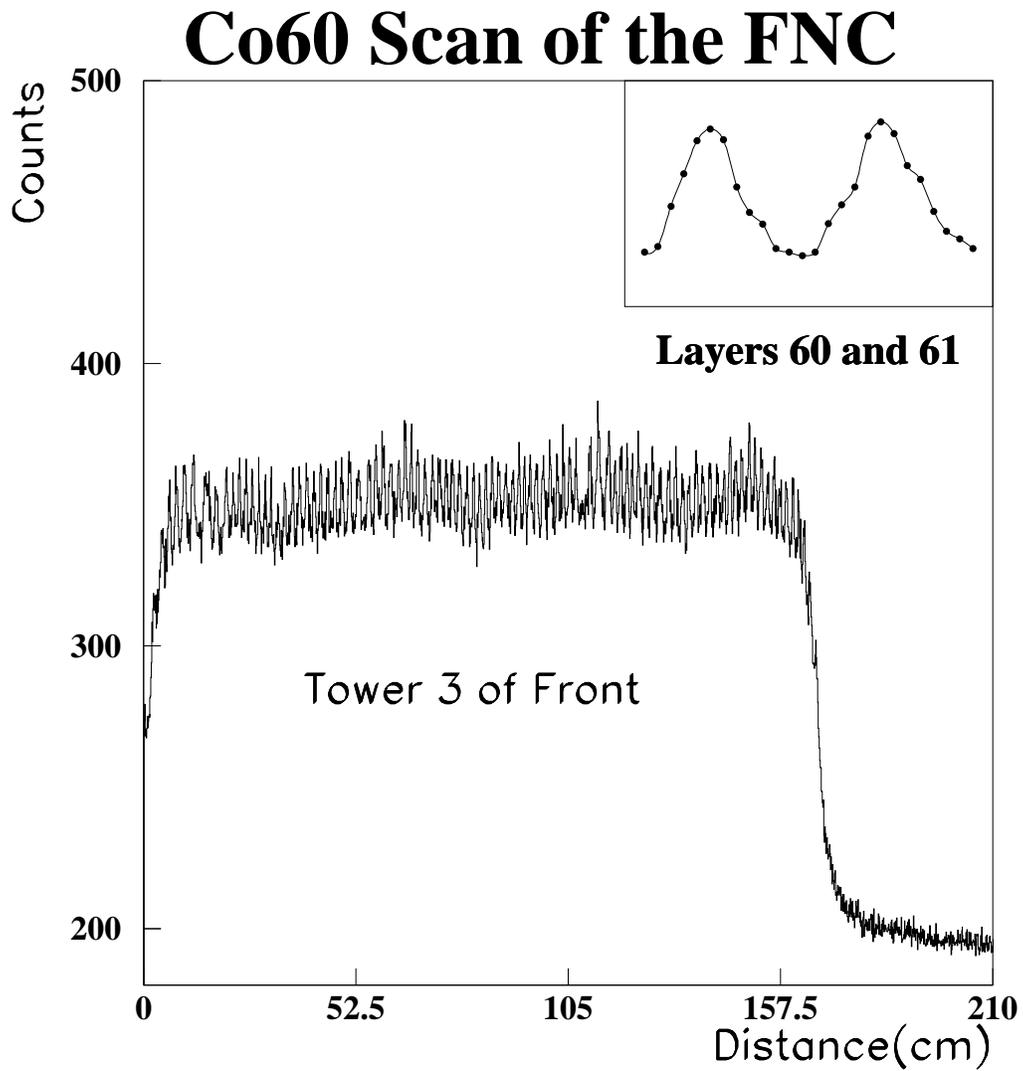}}
\caption{A far side $^{60}\rm Co$ source scan of tower 3 of the
front section of the FNC. The peaks of the 95 scintillator layers
are clearly visible. The inset shows an enlargement of a short
section.}
\label{front_scan}
\end{figure}

Figure~\ref{front_scan} shows a typical scan
of a front tower,
displaying the ADC output as a function of longitudinal
position for the PMT on the side of the calorimeter
opposite the source.
Each scintillator layer is visible as a peak.
As the source is withdrawn from the calorimeter,
the response falls and flattens
to become the pedestal which
must be subtracted to obtain the gain. 
The steepness of the falloff at the edge of
the calorimeter and the depth of the valley
between peaks indicate that the general 
response is a superposition of several scintillator
layers, as expected from the low energy (1.2 MeV) gamma rays
emitted by $^{60}$Co.

To measure the pedestal and scintillation signals, 
a scan is projected onto the ordinate of the plot.
The relative gain of a PMT, compared to all others,
is obtained by comparing
the fitted average peak position for its scan
after subtraction of the pedestal.
After correction for pedestal and relative gain, 
the projection of the peaks for all the scintillator layers
in all towers is distributed approximately as a Gaussian,
whose width to mean ratio (4.7\%) is a measure
of the uniformity of transmission of the optical
system, including scintillator and masked WLS.

\section{Test of the Calorimeter at CERN}
\label{cerntest}

Only the bottom part of the calorimeter was ready
for testing at CERN in the summer of 1994.
The calorimeter was installed in the H6 beam in the
North Area of the SPS. This beam
provided electrons, muons, pions and protons
up to an energy of 120 GeV. The positively charged
particle beam is mixed pions and protons with
no particle identification\cite{atherton}. Most of
the data was taken at 120 GeV.

The calorimeter was placed on a table whose 
horizontal and vertical position were adjustable
and measurable to $\pm 0.1$ cm.
A pair of crossed finger counters 
in front of the calorimeter
was used for triggering 
and to define the
location of the beam (Table~\ref{countertab}). 
Three large counters were
also positioned in front of the
calorimeter.
The signal from the last dynode of each of
the eight PMTs of the four 
central calorimeter towers was also
available for triggering on calorimeter
energy deposits.
\begin{table}[htbp!]
\begin{center}
\begin{tabular}{||c|c|c||} \hline\hline
Counter & Description & Coverage \\ \hline\hline
T1 & vertical finger counter & 1 cm x 0.5 cm \\ \hline
T2 & horizontal finger counter & 1 cm x 0.5 cm \\ \hline
T3 & & 20 cm x 20 cm \\ \hline
T4 & & 15 cm x 15 cm x 5 cm\\ \hline
T5 & muon counter & 20 cm x 20 cm\\ \hline
T6 & veto & 3 cm diameter hole\\ \hline\hline
\end{tabular}
\end{center}
\caption{Beam counters.}
\label{countertab}
\end{table}

The photomultipliers were read out with LeCroy 4300
charge integrating ADCs. The ADCs were gated so that
160 ns of the pulse from the calorimeter was
integrated. During data taking the ADC
pedestals were regularly measured, and the gains were
monitored with an internal test function in the units.
The gains were also checked daily using an external
charge injector.
Over the course of the beam test, which lasted one week,
individual ADC gains
varied less than 1 part per mil, and all gains were 
uniform to within 5 parts per mil.

\subsection{Coordinate System}

In the following we use a right-handed coordinate
system with $y$ vertically up and $z$ into the face
of the calorimeter.
The origin (0,0) is
taken to be 
between the two central towers 
halfway between the sides
(the center of the face of the calorimeter).
$X$ and $Y$ are measurements of $x$ and $y$
based on the table position;
$\xcal$ and $\ycal$, measurements 
based on energy deposits in the FNC.

\subsection{Measurement of Shower Position and Size}

The vertical position of a shower is determined 
using the centroid method by
taking an energy weighted average of the the tower
positions:
\begin{equation}
\ycal = \frac{\sum_{i=1}^{10} w_i y_i}{\sum_{i=1}^{10} w_i},
\label{eqn_y}
\end{equation}
where the sum is over towers in the front part of the
FNC, and the weights $w_i$ are functions
of the energy
deposits, $E_i$, in each tower. The weights are chosen to
minimize bias and to maximize resolution
in the position measurement.
The simplest 
method of estimating $y$ corresponds to choosing weights
which are linear in energy, $w_i=E_i$. Awes et al.\cite{awes}
have suggested that the weights be chosen to have instead
a logarithmic dependence on the energy deposits:
\begin{equation}
  w_i = \max \left[ 0, \ln \left( \frac{E_i}{f E_{\rm B}} \right) \right],
\label{eqn_w}
\end{equation}
where $E_{\rm B}= \sum_{i=1}^{10} E_i$,
and $0<f<1$ is a fractional cutoff. 
When the energy
deposited in a tower satisfies $E_i<f E_{\rm B}$, that
tower is not used in the determination of the vertical position.
The logarithmic dependence of the $w_i$ 
more closely matches the true transverse profile of energy
deposits in a hadronic shower than the linear dependence.
This estimator is simple, but it has an undetermined free
parameter $f$ which may be a function of the
deposited energy. 
In addition
the optimal value of the cutoff parameter may
depend on particle type, electron or
hadron. 

The vertical segmentation also allows us to estimate
the size of the shower using the RMS vertical shower
width:
\begin{equation}
W = \sqrt{
  \frac{\sum_{i=1}^{10} w_i y_i^2}{\sum_{i=1}^{10} w_i}
  - Y^2_{\rm CAL} }.
\label{eqn_yw}
\end{equation}
Since the width of electromagnetic showers 
is much less than
the width of hadronic showers,
this estimator enables
us to distinguish incident electrons from incident hadrons. 
The optimal weights for the width determination are not
necessarily the same as those for the position measurement.
Since we use the width $W$ to separate electromagnetic
from hadronic showers, it is most convenient to have an estimator that
does not depend on knowledge of the nature of the incident
particle. Otherwise an iterative procedure must be employed.
As we will discuss in more detail below, linear weights for
the shower size measurement give a 
hadron-electron separation which is close to that obtained 
with logarithmic weights whose cutoff parameter has been
optimized.

The horizontal position of a shower can be estimated from
the division of scintillation light between the left and
right sides of the calorimeter:
\begin{equation}
\xcal = \frac{\lambda}{2}\ln \left( \frac{L}{R} \right)
\label{eqn_x}
\end{equation}
where $\lambda$ is the effective attenuation length 
of the scintillator strips, and $L(R)$ is the charge
from the left(right) PMT.
If the light attenuation behaves as 
\begin{equation}
L,R\ \propto \exp (\pm a x + b x^2),
\label{eqn_atten}
\end{equation}
where $a$, $b$ are constants, 
and $x$ is the displacement
of the incident particle from the center of the tower,
then $\xcal$ is independent of $b$ and 
$\lambda=1/a$.

\subsection{Electrons}

The calorimeter was calibrated with
120 GeV electrons  
incident on the center of each tower. 
The high voltage for the tubes in the front was adjusted 
so that their mean response to 120 GeV electrons as measured in the 
Lecroy ADCs was 100 pC (400 counts). After lifting and rotating
the calorimeter by 180 degrees to make electrons incident 
on the rear towers,
the high voltage of each rear tube was set 
so that the response of the rear section to a muon passing
through the calorimeter would approximately equal the response
of the front section. 
Since the ratio of the number of
front layers to the number of 
rear layers is 95/39, the gain is considerably higher in the
rear; however, the energy deposits in the rear for
electrons and hadrons are small, so there is no saturation problem
for the very high energy particles at HERA.
Figure~\ref{energy}a shows the distribution 
of deposited energy, after calibration,
for 120 GeV electrons centered on towers 3 through 7, inclusive.
Pion contamination has been removed be requiring that
the vertical shower width be less 
than 3 cm (section~\ref{separation}).
The energy resolution for electrons centered on a tower
averaged 33\%/$\sqrt E $ independent of tower.
Electrons centered in the edge towers had
an energy response reduced by 3\% due to
transverse energy leakage.

\begin{figure}[htbp!]
\epsfysize=14cm
\centerline{\epsffile{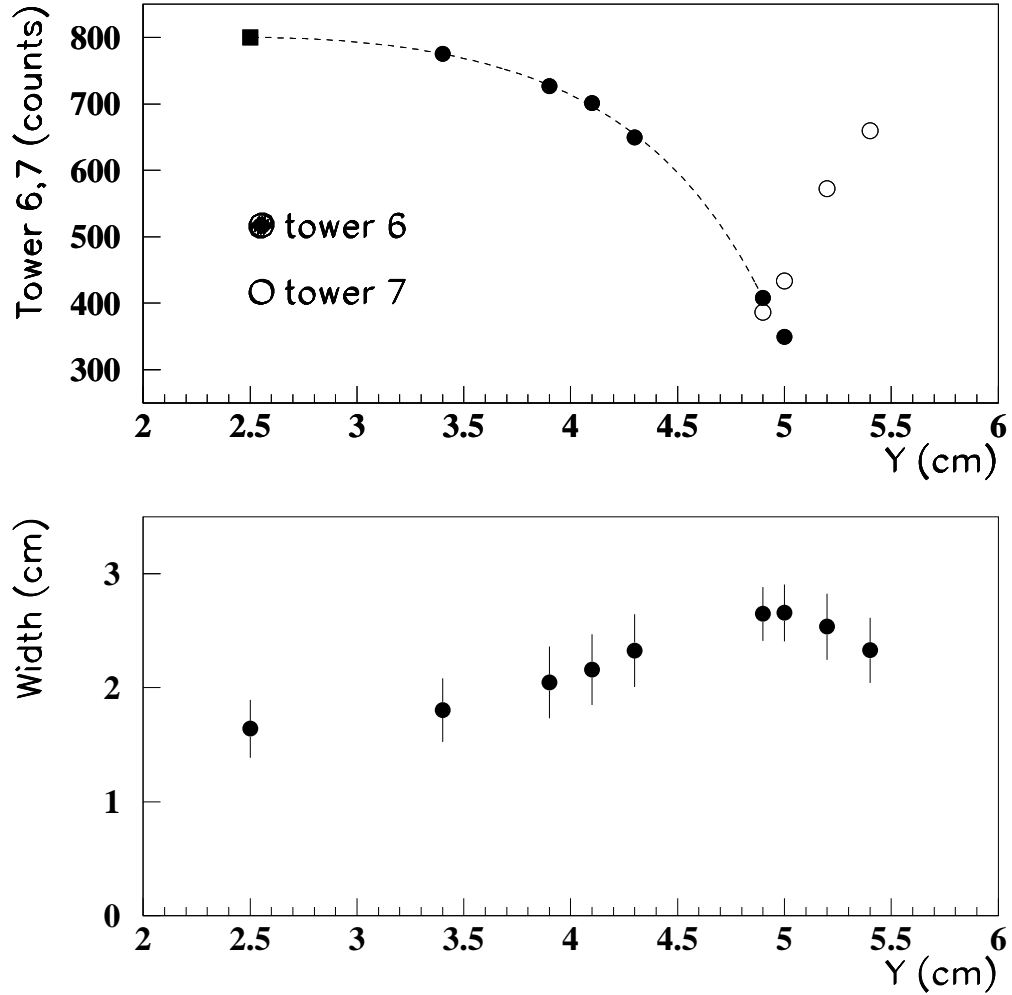}}
\caption{(a) The average energy, in ADC counts, deposited in 
in the central towers, 6 and 7,
as a function of $Y$ position for a vertical scan from tower 6
to tower 7. The point at $Y=2.5$ cm, the center of tower 6,
is fixed at 800 by the calibration with electrons.
The gap between towers is near $Y=5$ cm.
(b) The average width measured for 120 GeV electrons
in a vertical scan across the gap between the scintillator strips
of tower 6 and 7.
The data points are means; the error bars,
the corresponding RMS deviation of the distributions.}
\label{crack_width}
\end{figure}

The response of the calorimeter to electrons is dependent
on the position of the electron with respect to the gap
between the scintillator strips.
Figure~\ref{crack_width}a shows the energy deposited
in tower 6 when the electron beam is scanned from
tower 6 to tower 7.
The total
signal for electrons incident between two towers is 90\% of the
signal of an electron incident on the center of a tower. 
Moreover,
the energy resolution is degraded to 60\%/$\sqrt E$.
These effects are due
to the narrowness of electromagnetic showers
compared to the width of the region between strips.

\begin{figure}[htbp!]
\epsfysize=14cm
\centerline{\epsffile{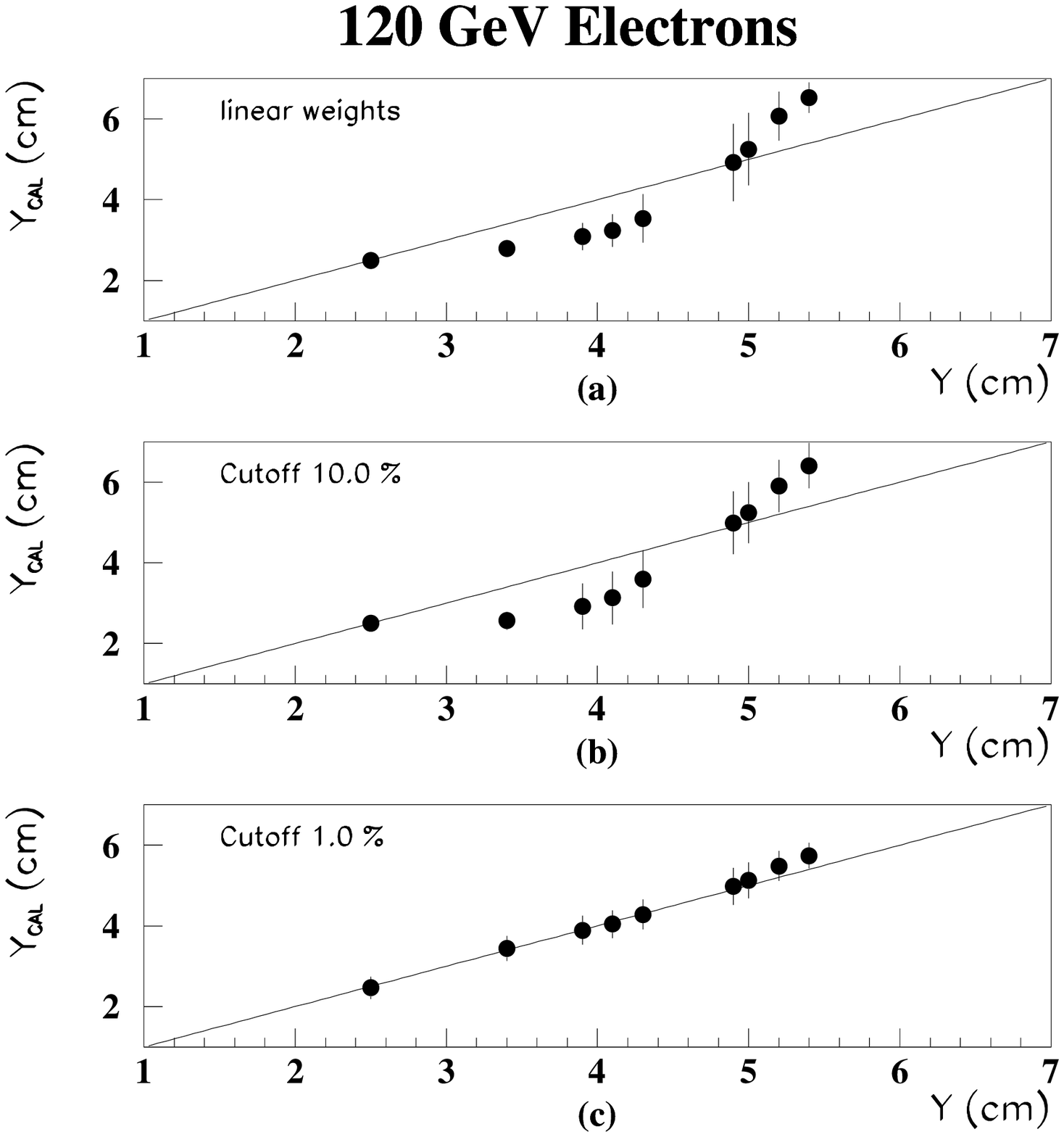}}
\caption{The average measured $\ycal$ as a function of $Y$
for 120 GeV electrons
in a vertical scan across the gap between scintillator strips.
Shown is $\ycal$ as measured using (a) linear weights and
using logarithmic weights with a cutoff of (b) 10\% and (b) 1\%.
The data points are means; the error bars the corresponding
RMS deviation of the distributions.}
\label{crack_pos}
\end{figure}

The small width of the shower 
compared with the strip width,
combined with the reduction in response at the tower
boundaries, is responsible for the large bias 
towards the tower center in the
measured electron vertical position
shown in Fig.~\ref{crack_pos}.
The bias grows with increasing cutoff parameter 
since for sufficiently large $f$
only the struck tower contributes to the sums in 
equations~\ref{eqn_y} and~\ref{eqn_yw}. 
The RMS deviation of the distribution also tends
to increase.
The variation with vertical position of
the shower width (linear weights) of electrons 
is shown in Fig.~\ref{crack_width}b, and is small
compared to the average width of hadronic showers
(see Section~\ref{separation}).

\begin{figure}[htbp!]
\epsfysize=14cm
\centerline{\epsffile{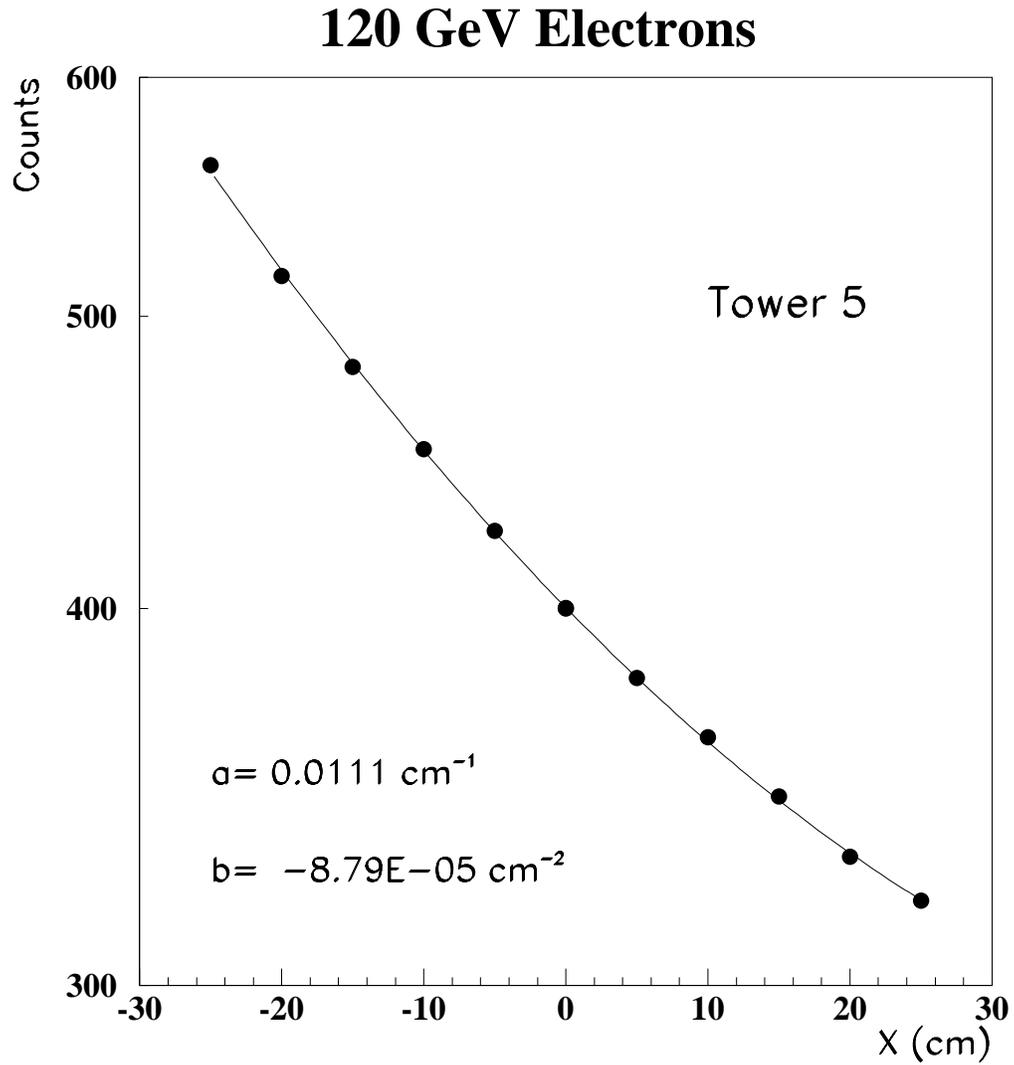}}
\caption{Horizontal scan of 120 GeV electrons along tower 5.
Charge from the near PMT (that towards which the beam
spot was moved) is plotted with $X<0$;
charge from the far (opposite) PMT is plotted with $X>0$.
Note the log scale on the $y$ axis.}
\label{attenl}
\end{figure}

Electrons were also used to determine
the effective attenuation length
for light in the scintillator strips. 
One half of each tower was scanned horizontally. For each
position the response of the right and the left photomultiplier
tubes was plotted as a function of the offset $x$
of the beam
from the center of the calorimeter, positive $x$ representing
an offset away from the PMT. 
Figure~\ref{attenl} shows the resulting attenuation curve for a typical
tower. The response is well represented by equation~\ref{eqn_atten}.
The tower-to-tower average value of
$a$ was found to be 1.08$\cdot 10^{-2}$ cm$^{-1}$
with a 3\% RMS spread, 
corresponding to an average effective
attenuation length of 92 cm; 
$b$ averaged 9.60$\cdot 10^{-5}$ cm$^{-2}$ with a 20\% RMS spread.

To compare the efficiencies of the wavelength shifting material 
in the front and rear section we measured
the number of photoelectrons collected
by the photomultiplier tubes, $n_{pe}$, which
is related to the width, $\sigma(L-R)$, of the 
difference distribution by
\begin{equation}
 \frac{\sigma (L-R)}{<L+R>} = \frac{1}{\sqrt n_{pe}},
\end{equation}
where $<L+R>$ is the mean of the sum.
By rotating the calorimeter 180 degrees
to make electrons incident
both front and rear,
we found that although in the rear we averaged 92 pe/GeV, in
the front we obtained only 26 pe/GeV.
This difference is due to different WLS lengths,
WLS material, and optical geometry
between front and rear, and it
worsens the $x$ position resolution by a factor
of $\sqrt 2$ compared to the estimate in section~\ref{description}.
Photostatistics makes only a
small contribution to the energy resolution since,
for 120 GeV electrons and 26 pe/GeV, the fluctuations are only 
about 2\%. 

\subsection{Hadrons}

\begin{figure}[htbp!]
\epsfysize=14cm
\centerline{\epsffile{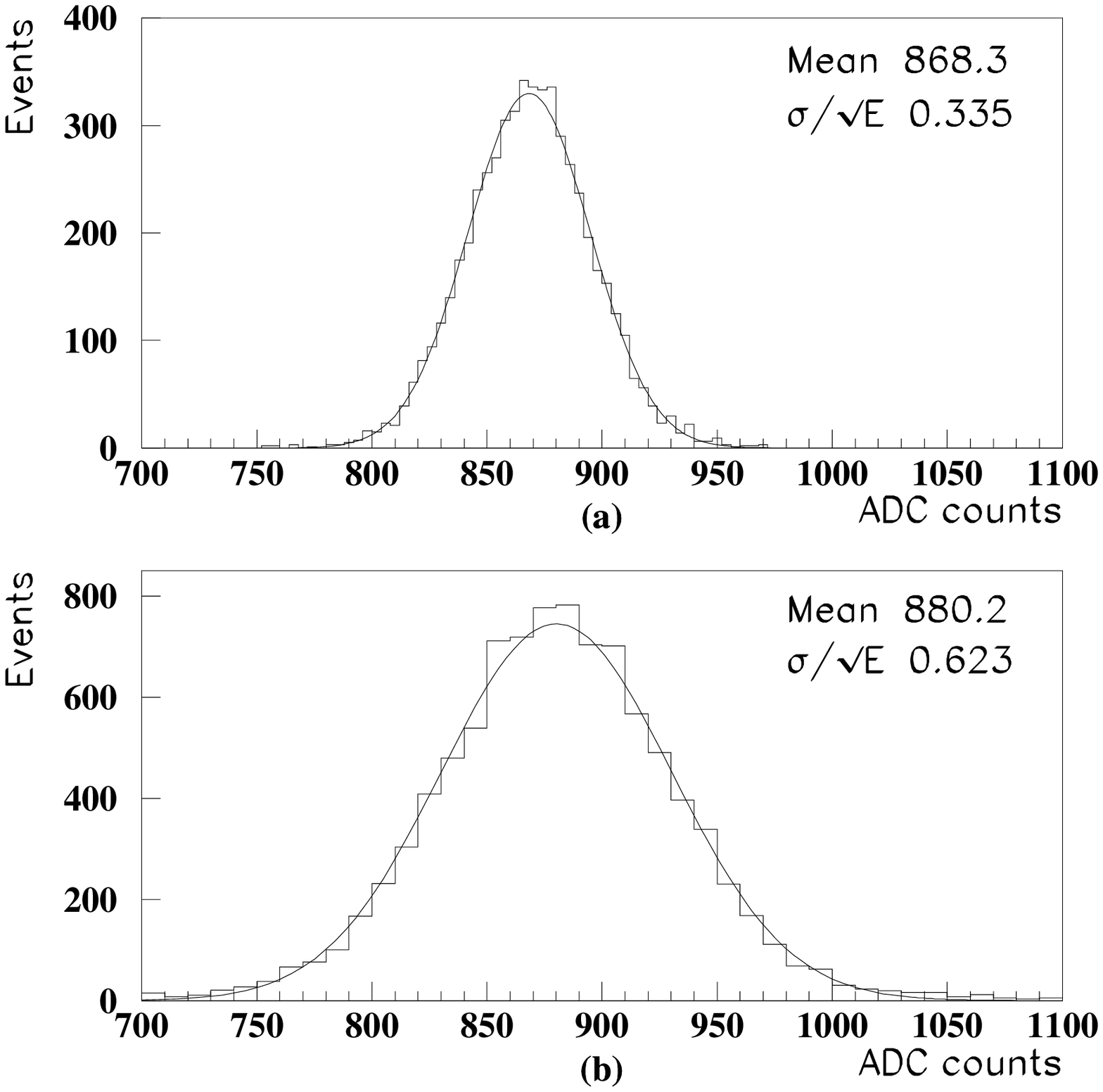}}
\caption{The energy distribution in ADC counts for 
(a) 120 GeV electrons centered on towers 3 to 7, 
and (b) 120 GeV pions centered on the calorimeter, 
between tower 5 and 6. Gaussians are fitted to
the distributions.
The pion contamination in the electron sample
has been removed by requiring $W<3$ cm (see text).}
\label{energy}
\end{figure}

\begin{figure}[htbp!]
\epsfysize=14cm
\centerline{\epsffile{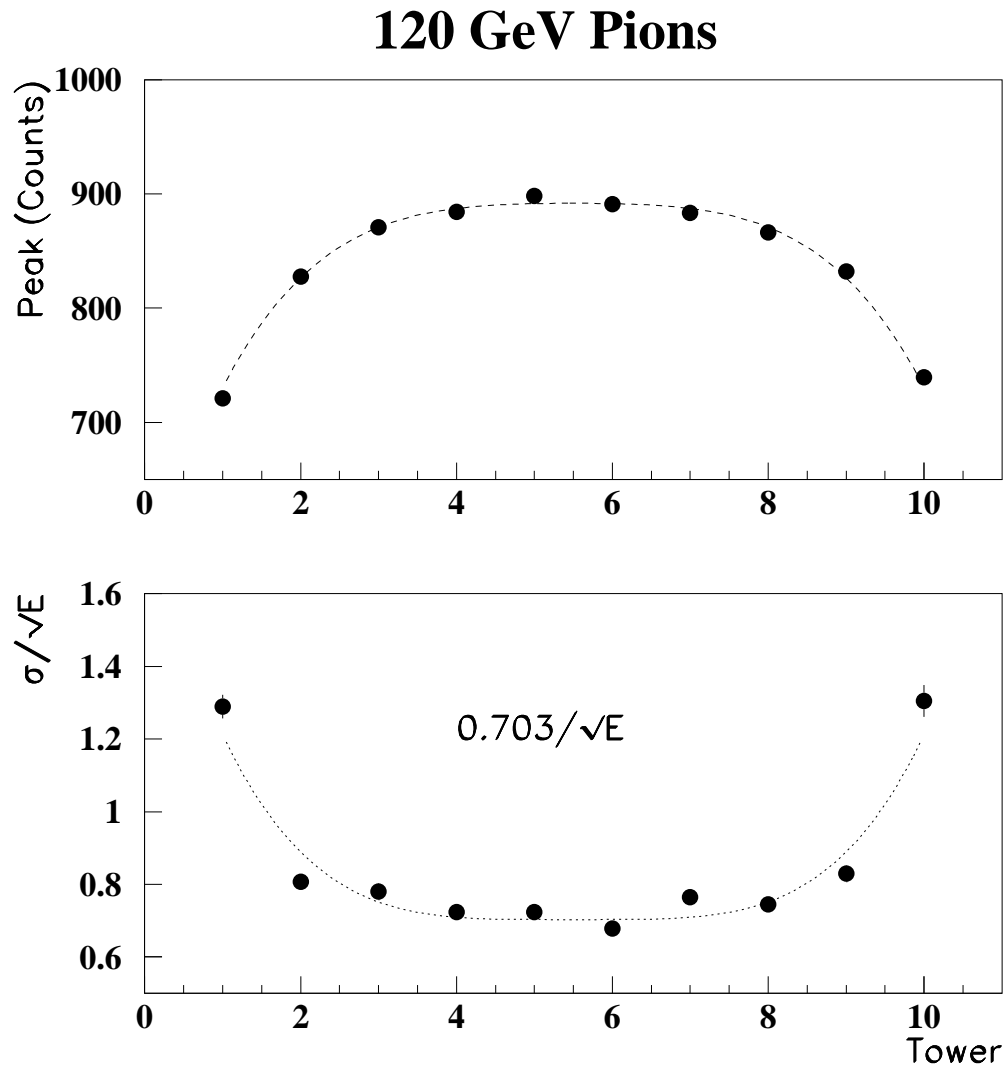}}
\caption{The energy response and resolution for a scan of the
calorimeter with 120 GeV pions in the center of each tower.
The curves are quadratic in $y^2$.}
\label{ptower}
\end{figure}

\begin{figure}[htbp!]
\epsfysize=14cm
\centerline{\epsffile{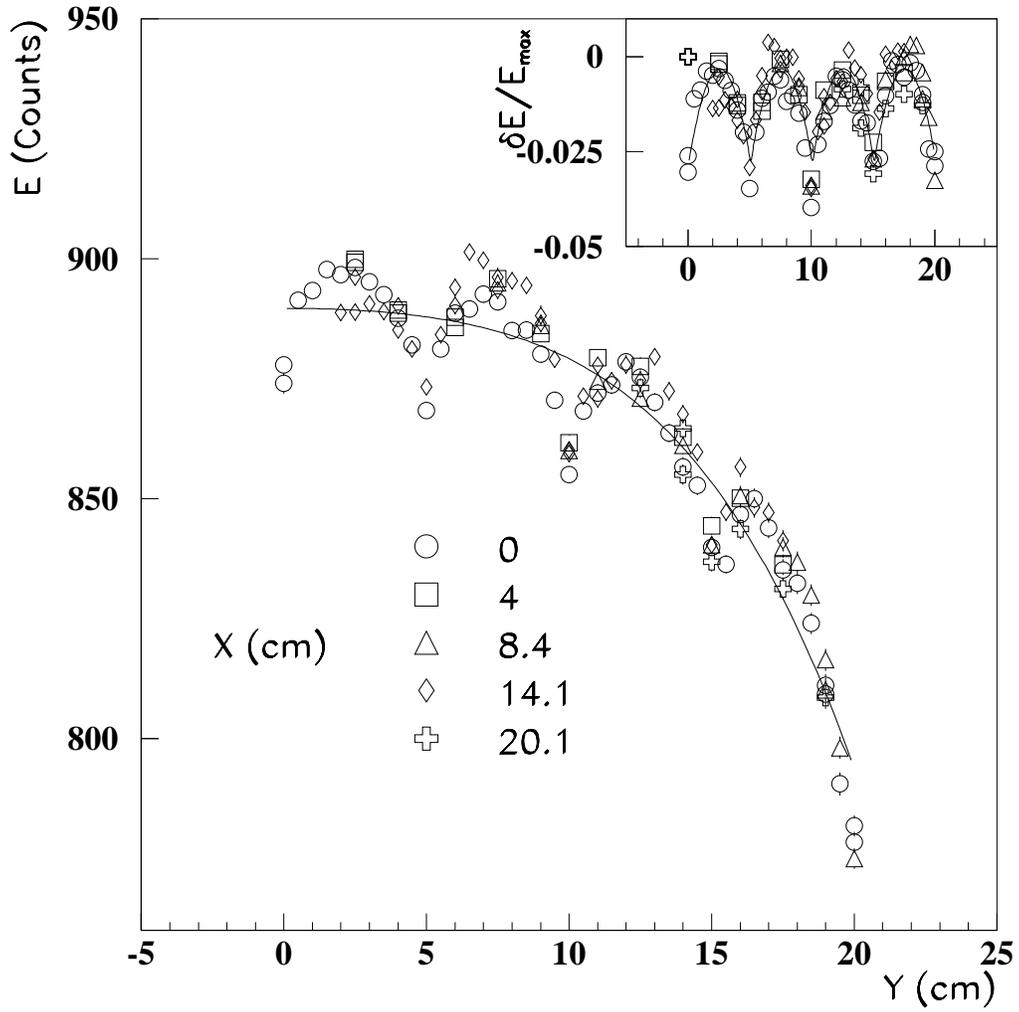}}
\caption{The measured energy as a function of $Y$ for
120 GeV pions incident on the face of the calorimeter
over a grid in $x$ and $y$. The energy response
has been corrected for the $x$ offset only.}
\label{eofy}
\end{figure}

Electron showers penetrate only about an interaction
length into the calorimeter. To refine the calibration
constants determined by electrons, 
120 GeV pions were
made incident on the center of each tower. Using these
data, correction factors were determined which equalized
the response of each PMT to hadrons. The correction factors
for hadrons were forced to average 1.0, and had an RMS spread of 5\%. 
The distribution of deposited energy for 120 GeV hadrons (pions and protons)
incident on the center of the calorimeter, between towers 5 and 6, is
shown in Fig.~\ref{energy}b.
For pions incident in the center of each tower,
the average response of the calorimeter
is shown in Fig.~\ref{ptower}.
The curves are quadratic in $y^2$. Leakage loss is important
whenever the beam is outside the four central towers.
When the beam is incident
in the center of the calorimeter, between 
towers 5 and 6, about 
2.5\% of the energy is lost compared to the case 
when the beam is in the center of tower 5 or 6
because of the gap between the scintillator strips
(see inset of Fig.~\ref{eofy}). 
Parameterized as $\sigma/\sqrt E$,
the energy resolution
for hadrons incident at the center of towers 5 and 6 is
0.70/$\sqrt E$ (Fig.~\ref{ptower}).
This improves to 0.62/$\sqrt E$
for hadrons incident at the center of the calorimeter,
between towers 5 and 6.

These results can be compared to the results for
electrons discussed above.
The electron to hadron ratio $e/h$ is 0.96 when the beams
are incident on the center of a tower, but it falls to 
about 0.9 when both are incident on the center of the
calorimeter in the gap between towers 5 and 6.
For electrons centered on a
tower, but hadrons incident on the center of the
calorimeter $e/h=0.99$ (see Fig.~\ref{energy}).
If the electron calibration constants
are used instead of the hadron calibration
constants, $e/h$ changes by 0.01.
As expected from the 
ratio of lead to scintillator thickness,
the calorimeter is 
over compensating.
This effect is magnified for electrons by the gap between
the scintillator strips; however, since compared
to the electrons the electromagnetic component 
of a hadronic shower is more spread
out transversely, we expect the `effective' $e/h$ to be higher.

In order to study energy leakage and position resolution 
for hadrons, the face of the calorimeter was scanned with
120 GeV pion over an $x,y$  grid.
Using this grid the effective attenuation length of the
scintillator was determined with hadronic showers.
The resulting
average attenuation length, 91.9 cm,
was in good agreement with the electron scan results. 

\begin{figure}[htbp!]
\epsfysize=14cm
\centerline{\epsffile{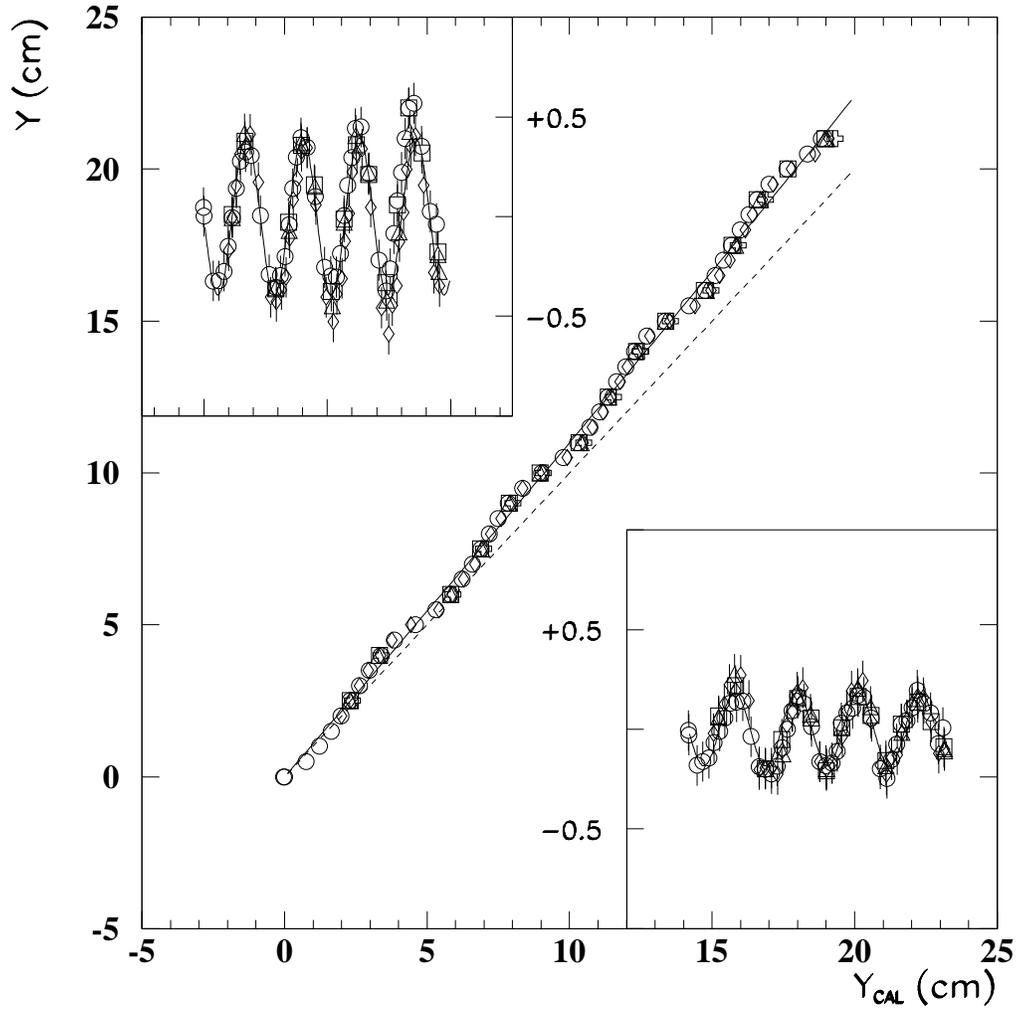}}
\caption{$Y$ as a function of $\ycal$, 
computed with linear weights,
for 120 GeV pions incident on the face of the calorimeter
over a grid in $x$ and $y$ (see Fig.~\ref{eofy} for the
grid and definition of symbols). The upper inset
shows the residuals from a quadratic fit. The lower
inset shows the residuals when $\ycal$ is 
computed with logarithmic weights (f=10\%). 
The dashed line is $Y=\ycal$.}
\label{ys}
\end{figure}

Figure~\ref{ys} shows the true vertical position 
of the calorimeter as a function of
its position determined using the
centroid algorithm with linear weights ($w_i=E_i$). 
The effect of the finite strip size is clearly
evident in the oscillation superimposed on an
average behavior. In addition, towards the edge of the
calorimeter the measured value falls below the true value
because of leakage. The average behavior can be
parameterized conveniently as 
$|Y|=1.080|\ycal|+0.002Y^2_{\rm CAL}$.
The residuals, shown in the upper inset,
follow an approximate sine curve
with period 5 cm, the tower height. 

\begin{figure}[htbp!]
\epsfysize=14cm
\centerline{\epsffile{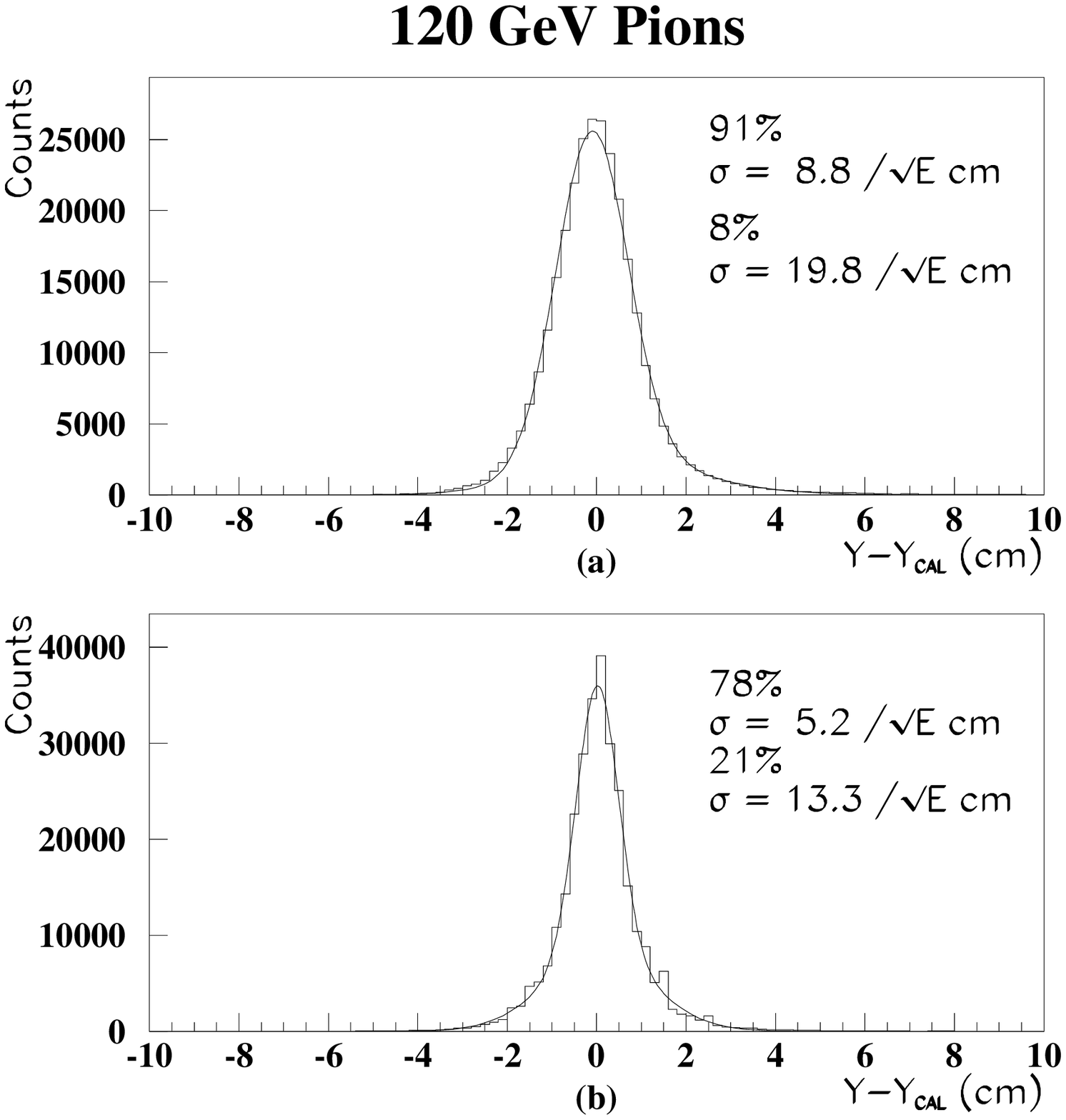}}
\caption{The distribution of residuals of the measured $y$ position
for 120 GeV pions incident on face of the calorimeter
a grid in $x$ and $y$ (see Fig.~\ref{eofy}) for (a) linear weights,
and (b) logarithmic weights (f=10\%). The distributions are
fit to the sum of two Gaussians.}
\label{yres}
\end{figure}

The overall vertical resolution
is determined by plotting
the $y$ residuals for the centroid algorithm
with the polynomial correction for all points of the grid.
The residual distribution is well
fit by a sum of two Gaussian (Fig.~\ref{yres}a). 
The largest component has
a weight of 0.91 and an RMS width of 0.8 cm for 120 GeV pions.
Because this width is larger than the size of the
oscillation, the $y$ position cannot be corrected
for the oscillation
on an event by event basis.

In contrast to linear weights,
logarithmic weights depend on an arbitrary cutoff
parameter $f$ which may be chosen 
to optimize the position resolution,
minimizing bias.
Residuals for logarithmic weights with $f=10\%$ are shown
in the lower inset in Fig.~\ref{ys} and 
in Fig.~\ref{yres}b.
The bias due to leakage has been
considerably reduced
($|Y|=0.99|\ycal|+0.001Y^2_{\rm CAL}$),
and the resolution improved,
compared with the use of linear weights.

\begin{figure}[htbp!]
\epsfysize=14cm
\centerline{\epsffile{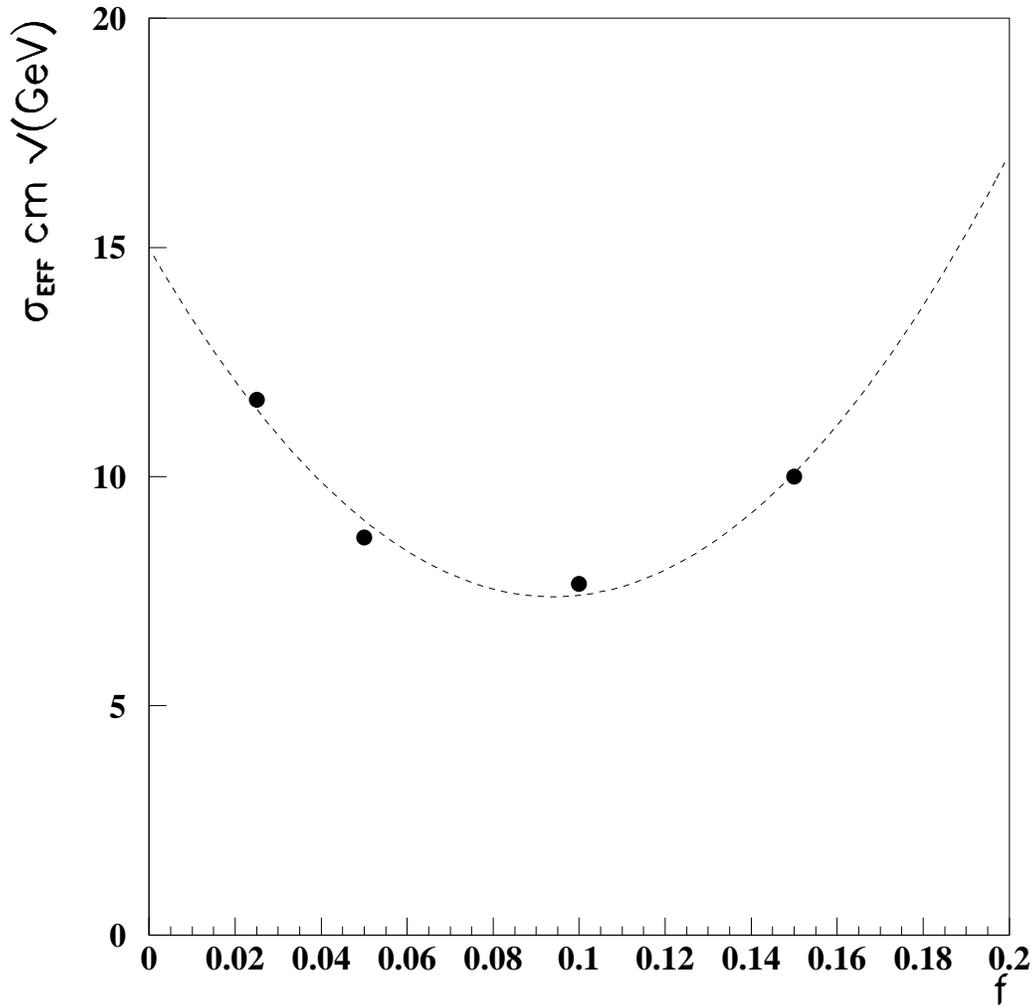}}
\caption{The variation of the effective vertical
position resolution, determined with logarithmic weights,
as a function of the cutoff fraction $f$. The minimum
of the curve, where the resolution is best, occurs
at $f=9.4\%$.}
\label{fval}
\end{figure}

A measure of the effective overall resolution is 
$\sigma_{\rm EFF}$
\begin{equation}
 \sigma^2_{\rm EFF} = f_1 \sigma_1^2+ f_2 \sigma_2^2
\end{equation}
where $\sigma_i^2$ is the variance of Gaussian $i$, and 
$f_i$ is its fractional contribution to the overall
distribution. The resulting $\sigma_{\rm EFF}$ is plotted
in Fig.~\ref{fval} as a function of $f$. The best
vertical position resolution for 120 GeV pions
is 8 cm/$\sqrt E$ for $f=9.4\%$,
where the curve has a minimum.

Like the $y$ residuals, the distribution of $x$ residuals
is also well represented by a sum of two Gaussians:
97\% with $\sigma = 20$ cm/$\sqrt E$, and
3\% with $\sigma= 60$ cm/$\sqrt E$,
giving $\sigma_{\rm EFF}= 22.3$ cm/$\sqrt E$.
The resolution in $x$ is much poorer than in $y$;
nevertheless,
it is close to the value expected since
for 1 GeV incident energy
\begin{equation}
\delta x = \lambda/\sqrt{n_{pe}} \approx 92\ {\rm cm}/\sqrt(26) = 18\ {\rm cm}.
\end{equation}

The energy response varies over the grid
as a result of two
competing effects: a) energy leakage, and b) light enhancement.
As the beam is moved towards the edge of the calorimeter,
energy is lost due to leakage,
and the measured signal tends to decrease.
On the other hand, as the beam moves horizontally away from the
center the light collected by the PMTs increases because of the
characteristic light attenuation of the scintillator strips:
\begin{equation}
L+R \propto \exp(bx^2)\left(\exp(-ax)+ \exp(ax)\right).
\end{equation}
The net result is that the signal increases for horizontal
displacements at fixed $y$. The
increase is approximately linear in $x$ with
\begin{equation}
L+R \propto \left[1+(0.002+0.00006|y|)|x|\right].
\end{equation}
Figure~\ref{eofy} shows the energy response as a function of y
for all grid values with the $x$ dependence removed. 
Energy loss due to the gap between the scintillator strips is clearly 
visible as is also loss due to leakage. The leakage
component has a quadratic dependence in $y^2$;
the loss due
to the gaps between strips
behaves as a superposed oscillation. 

\subsection{Electron-Hadron Separation}
\label{separation}

A calorimetric measurement with the capability of
distinguishing between electromagnetic 
and hadronic showers is important for physics studies.
Because electromagnetic showers are much narrower than
hadronic showers, the measurement of the shower width 
allows a separation of particles by type. 
The use of logarithmic weights for the determination of the
vertical shower position reduces the bias due to the finite
strip size and lateral leakage,
and improves the resolution; however, the 
optimal choice of the cutoff parameter depends on incident
particle type. At 120 GeV, $f$ should be chosen to be
about 10\% for hadrons,
but should be close to 0 (no cutoff) for electrons.

The choice of cutoff for the width measurement can be
chosen to optimize the separation of electromagnetic and
hadronic showers. To estimate the optimal value of 
$f$ we define 
\begin{equation}
Sep \equiv \frac{<W_h>-<W_e>}{\sqrt{S_h^2+S_e^2}}
\label{eqn_s}
\end{equation}
as a measure of the electron-hadron separation,
where $<W_{h,e}>$ and $S_{h,e}$ are the means and 
standard deviations, respectively, of the width distributions
for hadrons ($h$) and electrons ($e$). All are implicitly
functions of the cutoff $f$, so the value of $f$
can be chosen to maximize $Sep$. 

\begin{figure}[htbp!]
\epsfysize=14cm
\centerline{\epsffile{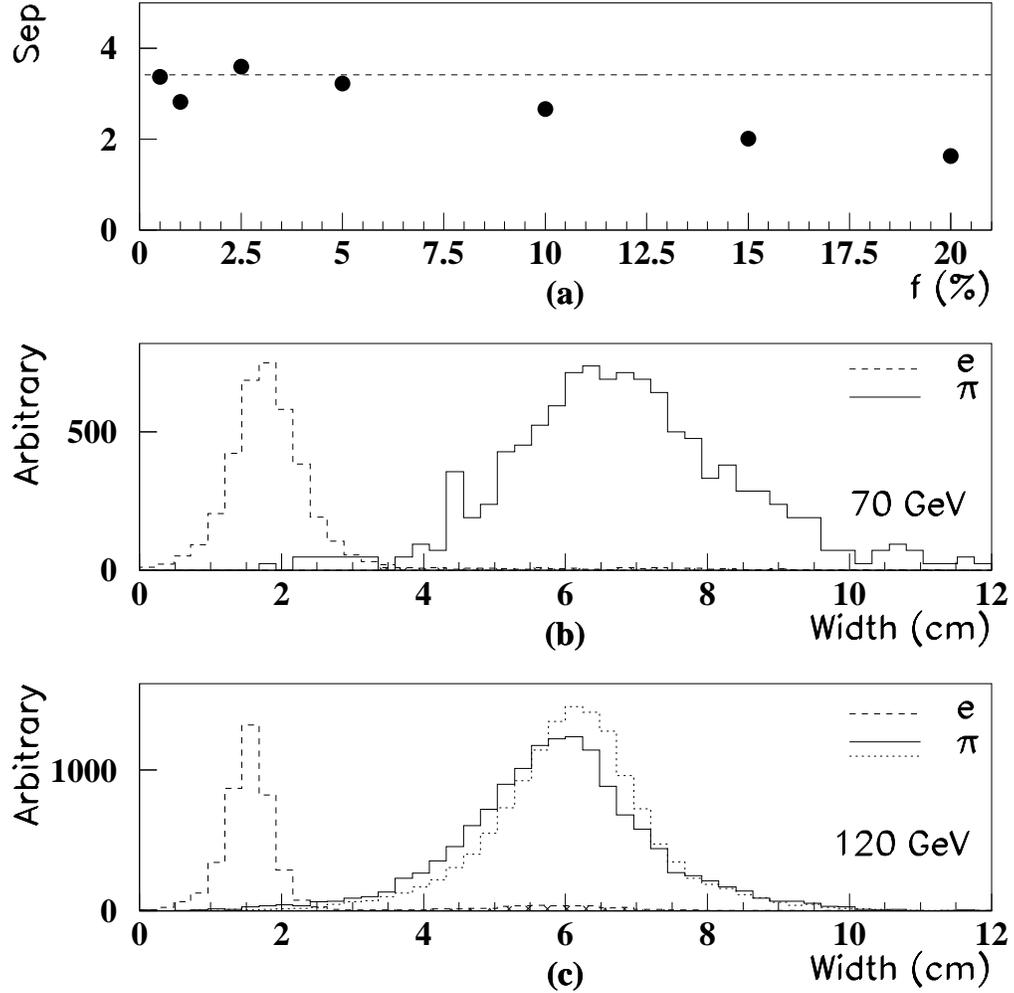}}
\caption{(a) The electron-hadron separation variable,
$Sep$, as a function of the logarithmic weight cutoff
parameter $f$. The dashed line shows the value of $Sep$
for linear weights. Below (a), the distribution of shower
widths, computed with linear weights, for (b) 70 GeV  
and (c) 120 GeV electrons and pions. 
The electrons are incident on the center of towers.
The pions are incident at the center of a tower
(full) and at the center of the calorimeter (dotted).
The pion contamination in the electron sample is visible.}
\label{ehsep}
\end{figure}

The dependence of $Sep$
on $f$ for 120 GeV electrons and hadrons is shown in 
Fig.~\ref{ehsep}. $Sep$ is insensitive to $f$ at low
values of $f$, but 
falls approximately linearly for $f\gap 5$\%.
Linear weights give the result shown by the
dashed line, $Sep=3.4$, 
close to best values obtained with logarithmic
weights, 
and have the advantage of being 
independent of $f$, which may depend on the incident
energy. As a result, we use linear weights for
measurement of shower widths instead of the logarithmic
weights chosen for the shower centroid measurement.

\begin{figure}[htbp!]
\epsfysize=14cm
\centerline{\epsffile{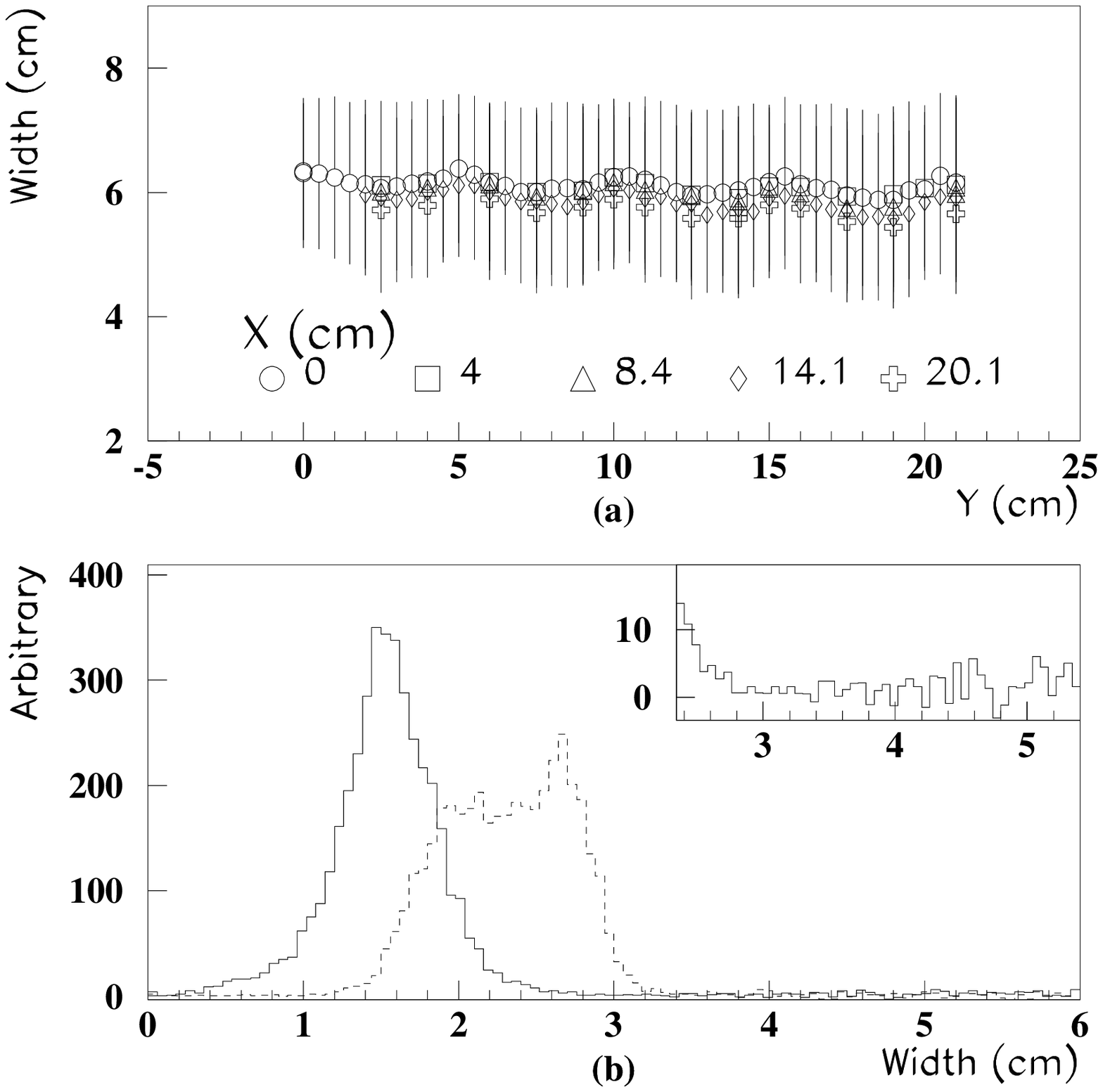}}
\caption{(a) The variation of the mean and RMS of 
the width distribution of 120 GeV pions with
vertical and horizontal position in the calorimeter.
The data points are the mean of the distribution;
the `error' bars the RMS.
(b) The distribution of shower widths, 
computed with linear weights and 
after subtraction of the pion contamination,
for 120 GeV electrons centered on a strip (full), and 
a scan of electrons starting from 0.9 cm above the
center of tower 6 to tower 7 (dashed, see 
Figs.~\ref{crack_width} and~\ref{crack_pos}). The inset shows
the tail of the distribution for centered electrons.}
\label{ywidthy}
\end{figure}

The vertical width distribution is plotted in 
Fig.~\ref{ehsep}b and c
for 70 and 120 GeV electrons and 
pions incident on towers 2 through 9. 
Although the distributions do not change much with energy,
there is a small reduction in spread
and decrease in mean. 
Figures~\ref{ehsep}c and~\ref{ywidthy}a
show that mean and standard deviation of the shower width 
distribution for hadrons does not change significantly with 
the incident vertical position.
A $W$ cut chosen in the range 2-4 cm will clearly
separate electrons from hadrons. 
Only 2.3\% of hadrons incident at the center 
of the calorimeter will be
lost with a $W$ cut of 3.5 cm; 
the loss rises to 4.0\% for hadrons
incident at the center of a strip (Fig.~\ref{ehsep}).  

The same analysis for electrons is complicated
by pion contamination in the electron sample, which is 
shown as a broken line in Fig.~\ref{ehsep}.
To subtract the pion contamination, we normalize 
the width distribution for pions 
to the raw (unsubtracted) distribution for electrons
in the region $W>6$ cm. The
result is shown in Fig.~\ref{ywidthy}b for electrons
incident in the center of a tower, and electrons scanned
vertically from tower 6 to 7. The inset shows the tail of
the distribution for centered electrons.
About 1.5\% of the electron showers have a width greater
than 3.5.

\subsection{Muons}

\begin{figure}[htbp!]
\epsfysize=14cm
\centerline{\epsffile{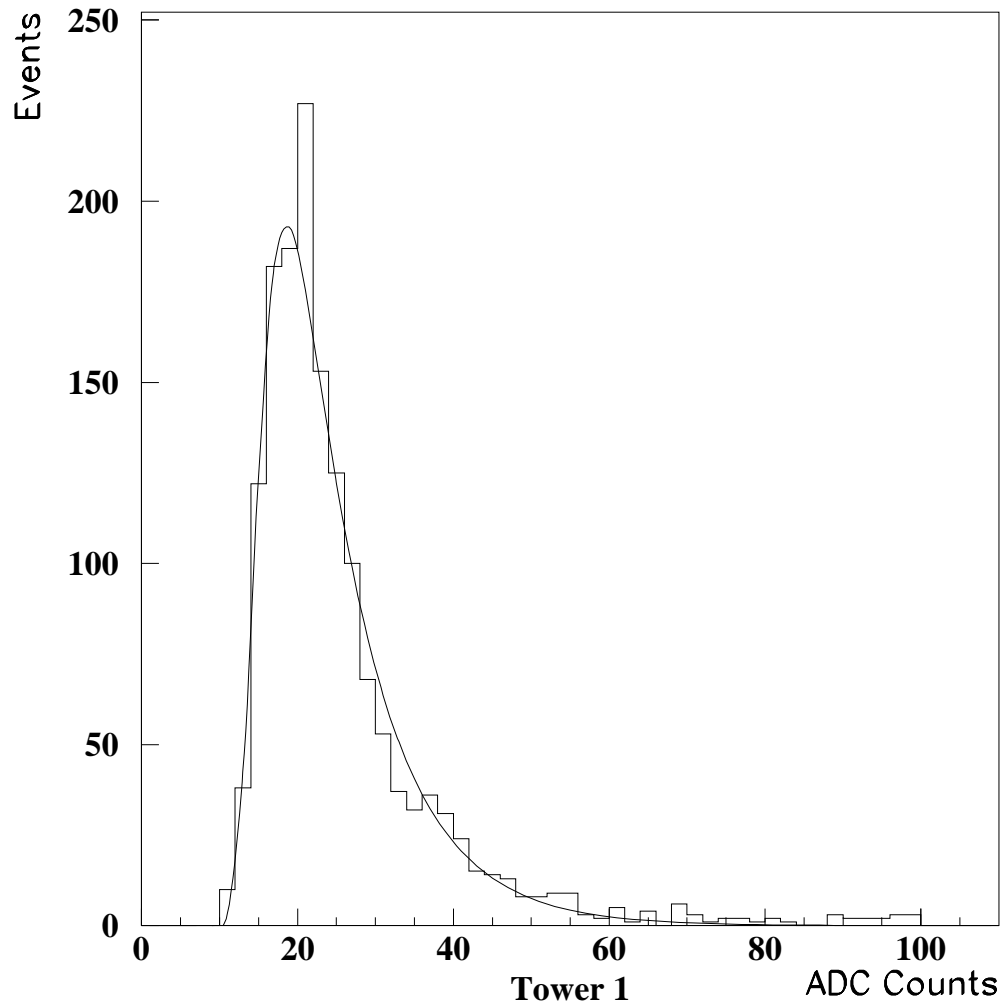}}
\caption{The energy deposited by a 120 GeV muon in the front
part of the calorimeter.}
\label{muon}
\end{figure}

Although the signal from penetrating muons is low because
of their small energy deposits, they can be useful for calibrating
and monitoring the FNC in HERA using either cosmic
rays or beam halo muons.
To study the response of the calorimeter to high energy muons,
a special trigger was implemented for beam particles which
penetrated the calorimeter.
Figure~\ref{muon} shows a typical
energy distribution for a muon in the bottom front tower.
The peak position $x_0$ is determined
by fitting the distribution to a convenient
functional form:
\begin{eqnarray}
m(x) &=&A\exp \left(a\left(y-\exp(-y)\right)\right), \\
 y   &=&\frac{x-x_0}{\sigma} \nonumber
\end{eqnarray}
and $A$, $a$, $x_0$, and $\sigma$ are parameters.
When $a=1/2$, $m(x)$ is the Moyal function\cite{moyal},
an analytic approximation for the Landau distribution.
The tower-to-tower spread of the peak position
is 4\%. 

\subsection{Neutrons}

The absolute energy scale is best determined by making
a beam of known energy incident on the calorimeter. 
If deuterons were to be accelerated at HERA, nuclear 
stripping reactions would make neutrons of known
energy incident on the calorimeter. At present
only protons are accelerated,
and the absolute energy scale must be determined using
beam gas interactions which produce leading neutrons by
charge exchange. If deuterons were accelerated at
HERA in order to test
the feasibility of using charge exchange
reactions to determine the absolute energy scale, a short
experiment was performed.

\begin{figure}[htbp!]
\epsfysize=14cm
\centerline{\psfig{file=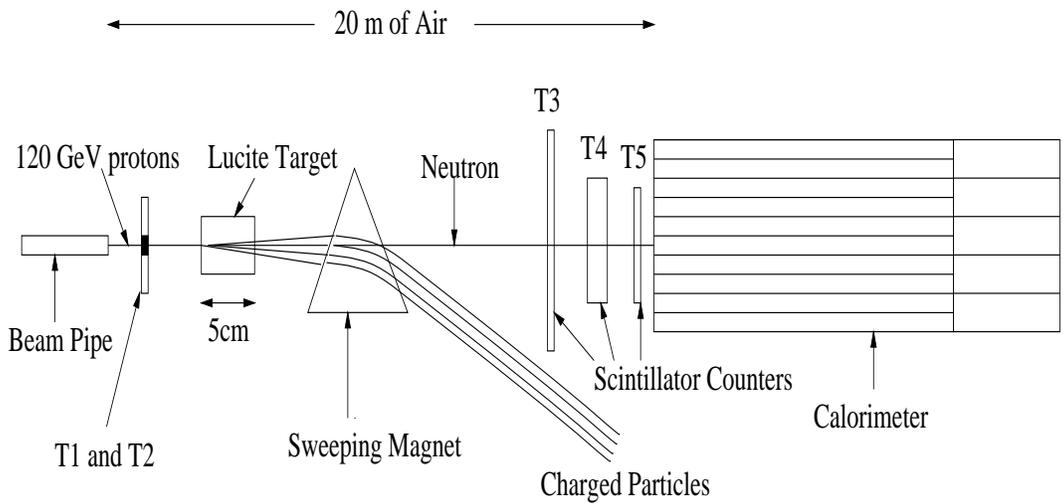,width=14cm,height=6.5cm,angle=-90}}
\caption{A schematic of the FNC test setup at CERN.}
\label{cern_tb}
\end{figure}

Figure~\ref{cern_tb} shows the experimental arrangement. A
lucite target was placed in the beam
20~m in front of the calorimeter.
Immediately after the target a sweeping magnet was tuned
to deflect beam particles downward. With the magnet tuned
to full field beam particles ( protons and pions )
were deflected downwards by
11 cm at the calorimeter. It was necessary to devise
a trigger that ensured that only events arising from
the impact of a neutral particle in the central part of
the calorimeter were accepted.

Crossed finger counters in front of the target 
defined the beam position
(see Table~\ref{countertab} and Fig.~\ref{cern_tb}). 
The trigger required
a coincidence of these two counters and energy deposited
in the center of the calorimeter
For the trigger,
the deposited energy was measured by summing and
discriminating the signals from the last dynode
of the PMTs of the four central towers.
In order to veto charged particles,
three scintillation counters were placed directly in front
of the calorimeter.
In addition, only showers near the center of the calorimeter 
were considered
in order to reduce energy leakage. 
In summary the data were required to satisfy:
\begin{eqnarray}
 ({\rm T}1 > {\rm mip})\cdot ({\rm T}2 > {\rm mip})\\
 ({\rm T}3 < {\rm mip})\cdot ({\rm T}4 < {\rm mip})\cdot ({\rm T6} < {\rm mip})\\
 {\Sigma E}_{\rm dynodes} > {\rm threshold}\\
 R < 5 \ {\rm cm},
\end{eqnarray}
where
\begin{equation}
  R = \sqrt{X^2_{\rm CAL} + Y^2_{\rm CAL}}.
\end{equation}
The shower position $\xcal$
was determined by light division;
$\ycal$, by
the centroid method with logarithmic weights.

\begin{figure}[htbp!]
\epsfysize=14cm
\centerline{\epsffile{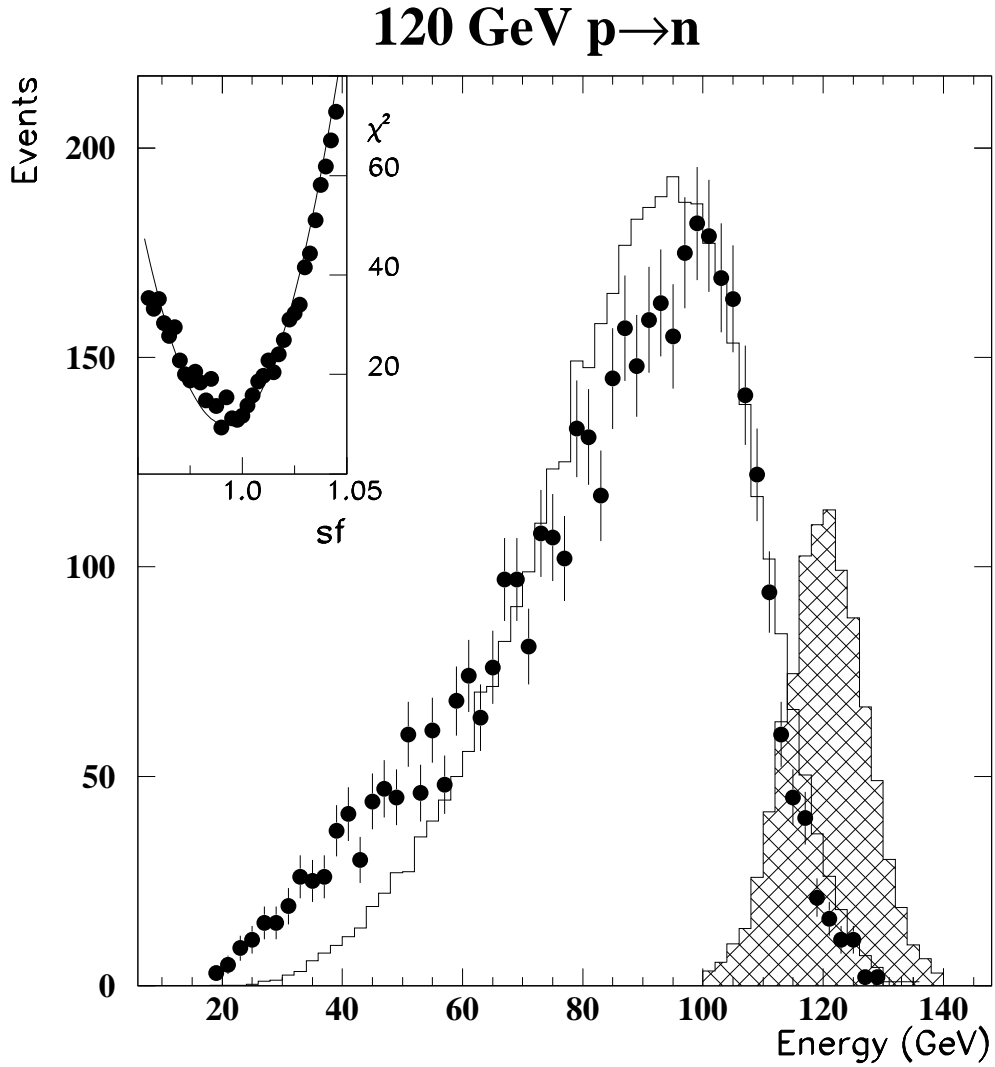}}
\caption{The neutron energy spectrum
(solid points) compared to the
spectrum predicted by one pion exchange (open histogram).
The hatched histogram shows the measured energy
spectrum for 120 GeV hadrons (see Fig.~\ref{energy}b).
The inset shows the $\chi^2$ comparison of the measured neutron
spectrum and the predicted neutron spectrum as a function
of the neutron to beam hadron energy scale factor (sf).}
\label{neutron}
\end{figure}

The observed energy distribution is shown in
Fig.~\ref{neutron}.
In the same figure, the hatched histogram shows
the response  of the calorimeter to 120 GeV hadrons.
The mean value of the latter distribution
determines the energy scale for the figure.

The one pion exchange (OPE) model of the reaction 
$p\rightarrow n$ predicts that the
cross section, integrated over scattering angles, is given by
\begin{equation}
\frac{d\sigma }{dE}=f_{\pi /p}(E)
      \sigma_{\pi}(\Delta E),
\label{eqn_ope}
\end{equation}
where $f_{\pi /p}$ is the flux of virtual pions 
associated with the incoming proton;  
$\sigma_{\pi }$ is the total pion-target
cross section  
for a virtual pion beam energy of $\Delta E=E_b-E$.
Any dependence of $\sigma_{\pi}$
on the virtuality of the
exchanged pion has been ignored in
equation~\ref{eqn_ope}, so 
$\sigma_{\pi}$ can be estimated using
a parameterization
($\Delta E>4$ GeV), or an interpolation table
($\Delta E<4$ GeV)\cite{pdg}.
Several forms have been postulated
for the flux factor $f_{\pi /p}$
in theoretical and 
phenomenological studies of OPE.
They fall into two classes: Regge\cite{bishari,lf,kopeliovich}
and light cone\cite{holtmann}.

Accordingly, to compare data and theory,
we extended the HERWIG\cite{herwig} Monte Carlo program
to generate events according to OPE folded with the
measured experimental resolutions.  
The predicted energy distribution, for the
light cone flux factor\cite{holtmann},
adjusted for resolution and acceptance is shown as the open
histogram in Fig.~\ref{neutron}.  
The OPE prediction is normalized to the data 
by the number of events.
As can be seen,
agreement is good for energies larger than 100 GeV.

The energy scale was varied and then compared with the
OPE prediction for $E>100$ GeV by computing the $\chi^2$ of
the difference distribution.  The variation of $\chi^2$ with the
scale factor is shown in the inset to Fig.~\ref{neutron}. The minimum
occurs for a scale factor of 0.985. If the Regge\cite{lf,fnc2}
form for the flux factor is used instead, the same procedure 
yields a minimum at 1.003, in complete agreement with the incident
proton beam; in this case, however, the minimum
$\chi^2$ is worse (17 rather than 10).
We conclude that using charge
exchange neutrons the absolute energy scale can be determined
to better than 1.5\% with the chief source of error being
the theoretical uncertainty on the form of
the virtual pion flux factor.

\section{Monte Carlo Studies of the FNC}

Since the spring of 1995, the FNC has been operating in
the ZEUS experiment at HERA where it is used in the study
of leading neutrons produced at small angles with energies
$E \gap$ 100 GeV. Because the calorimeter has been calibrated
and tested only in beams of energy up to 120 GeV,
and because the top was not present for the beam tests, 
we must rely on a Monte Carlo simulation to predict
the response
of the calorimeter to high energy particles.
We have modeled the FNC using the GEANT 3.13\cite{brun} program,
upon which the simulation of the full ZEUS detector is based.
In this section we present some results from the simulation
which can be compared to our test beam data.

\begin{figure}
\epsfysize=14cm
\centerline{\epsffile{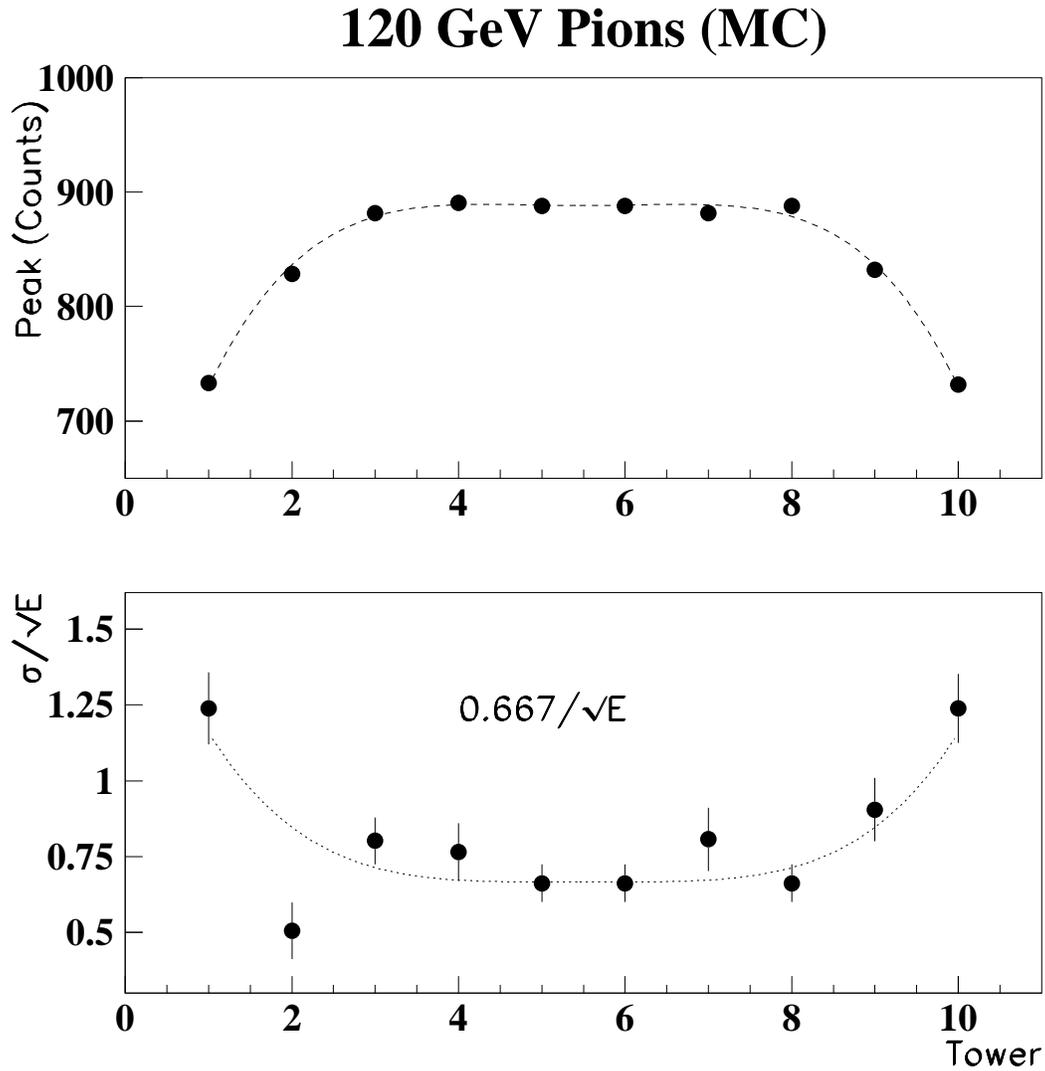}}
\caption{Monte Carlo simulations of the 
response of the calorimeter to 120 GeV pions
incident on the center of each tower. The energy resolution,
due to shower fluctuations, that is, without photostatistics,
is shown in the lower graph. The corresponding 
test beam data are shown in Fig.~\ref{ptower}.}
\label{mc_eres}
\end{figure}

For 120 GeV electrons and pions incident on the center
of towers 5 and 6 the GEANT
simulation predicts an electron to hadron response
ratio of 0.98, in agreement with
the measured value of 0.96.
The simulated response to pions incident on 
the center of each tower is shown in Fig.~\ref{mc_eres}.
The energy loss due to leakage when the
beam is incident near the edge of the calorimeter is 
also in
agreement with the data as is the degradation of the
energy resolution
(compare with the data shown in Fig.~\ref{ptower}).

The Monte Carlo gives an
energy resolution due to shower fluctuations alone,
that is, neglecting fluctuations due to photostatistics, 
of 0.66/$\sqrt E$ for hadrons incident
at the center of towers 5 and 6.
We have also
used the Monte Carlo to predict, at higher energies, 
the overall energy response of the calorimeter and its
expected energy resolution due to shower fluctuations. 
The Monte Carlo predicts that
for neutrons
centered on the calorimeter without the top,
the energy response
is linear up 800 GeV.
Fitting the energy resolution as a function
of incident energy, we find 
\begin{equation}
\frac{\delta E}{E}=\frac{0.54}{\sqrt{E}}\oplus 0.03
\end{equation}
where $E$ is in GeV. If fluctuations due to
photostatistics are added, the sampling constant 0.54 
will increase to 0.58. The top section is present
at HERA, but the neutrons are predominantly incident
on towers 7 and 8. 

\begin{figure}[htbp!]
\epsfysize=14cm
\centerline{\epsffile{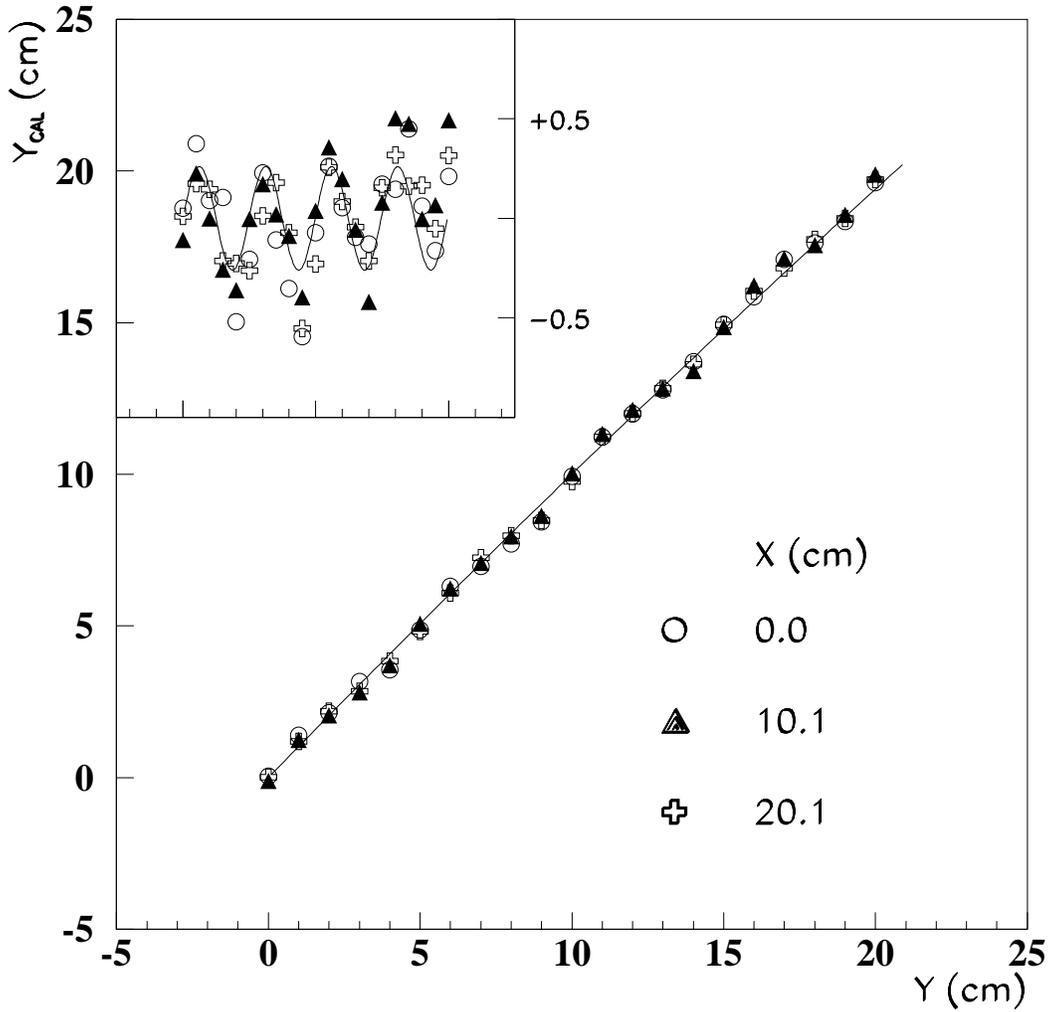}}
\caption{Monte Carlo simulation of $\ycal$, determined
by logarithmic weights with f=10\%, as a function of
$Y$ for 120 GeV pions incident on the calorimeter
over a grid in $x$ and $y$.
See Fig.~\ref{ys} for the corresponding plot made with
test beam data.}
\label{mc_ys}
\end{figure}

Just as for data, 120 GeV pions were studied with a grid
over the face of the calorimeter. Fig.~\ref{mc_ys} shows the
predicted dependence of $\ycal$ on $Y$, the vertical
impact position, for three values of horizontal impact position,
$x= 0$, 10.1, and 20.1 cm. $\ycal$ is calculated
with logarithmic weights and a cutoff parameter of f=10\%.
The simulation shows 
behavior and biases similar to 
the data (Fig.~\ref{ys}, lower inset); 
in particular, 
the value of $\ycal$ is
biased towards the nearest tower center.
The $Y$ residual distribution is Gaussian with
mean 0 and width 7.3 cm/$\sqrt E$ (compare
to the data width of 8.0 cm/$\sqrt E$).
The $X$ residual distribution
is Gaussian with mean 0 and width 10.3 cm/$\sqrt E$. 
This width is due to transverse shower fluctuations.
If photostatistics fluctuations
are also included, the width constant increases to 20.7 cm.
This should be compared with the data value of 22.3 cm.

\begin{figure}[htbp!]
\epsfysize=14cm
\centerline{\epsffile{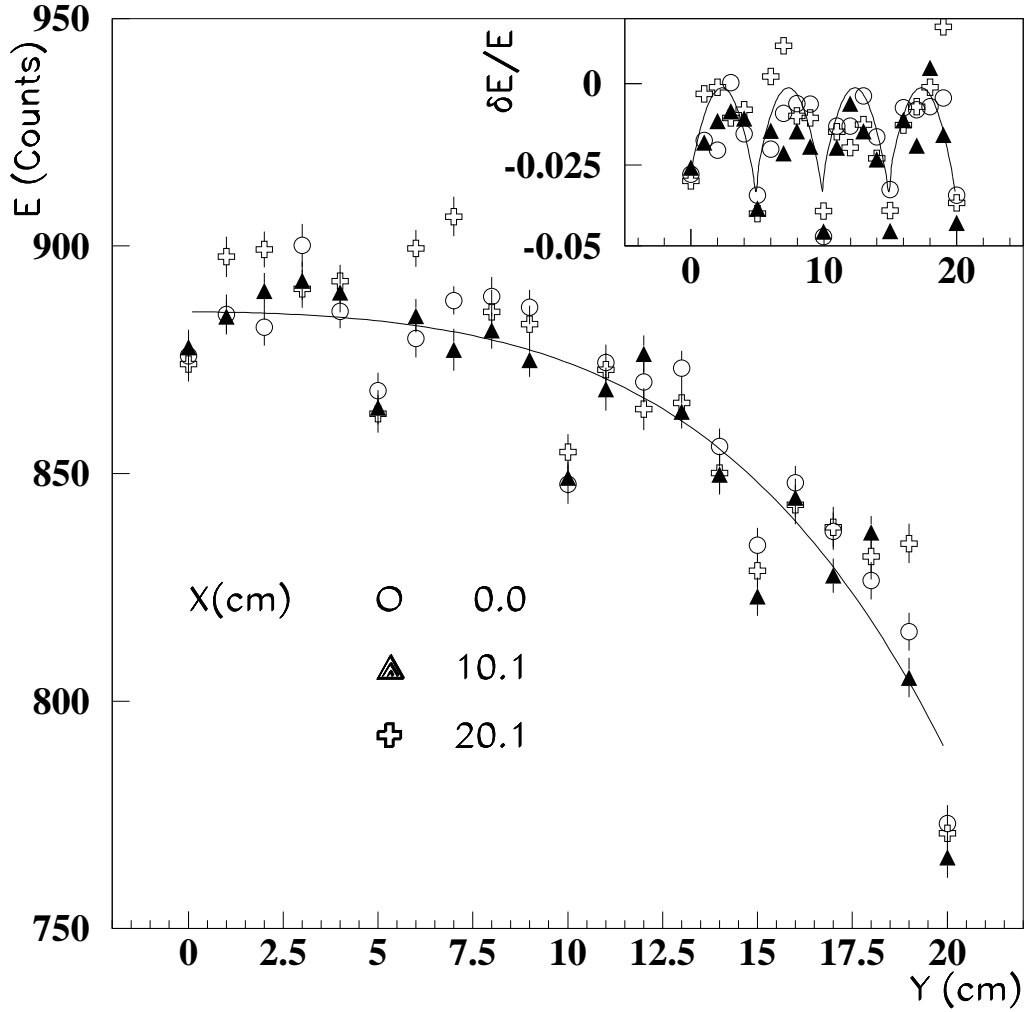}}
\caption{Monte Carlo simulation of the
measured energy as a function of $Y$ for
120 GeV pions incident on the face of the calorimeter.
See Fig.~\ref{eofy} for the corresponding plot made
with test beam data.}
\label{mc_eofy}
\end{figure}

\begin{figure}[htbp!]
\epsfysize=14cm
\centerline{\epsffile{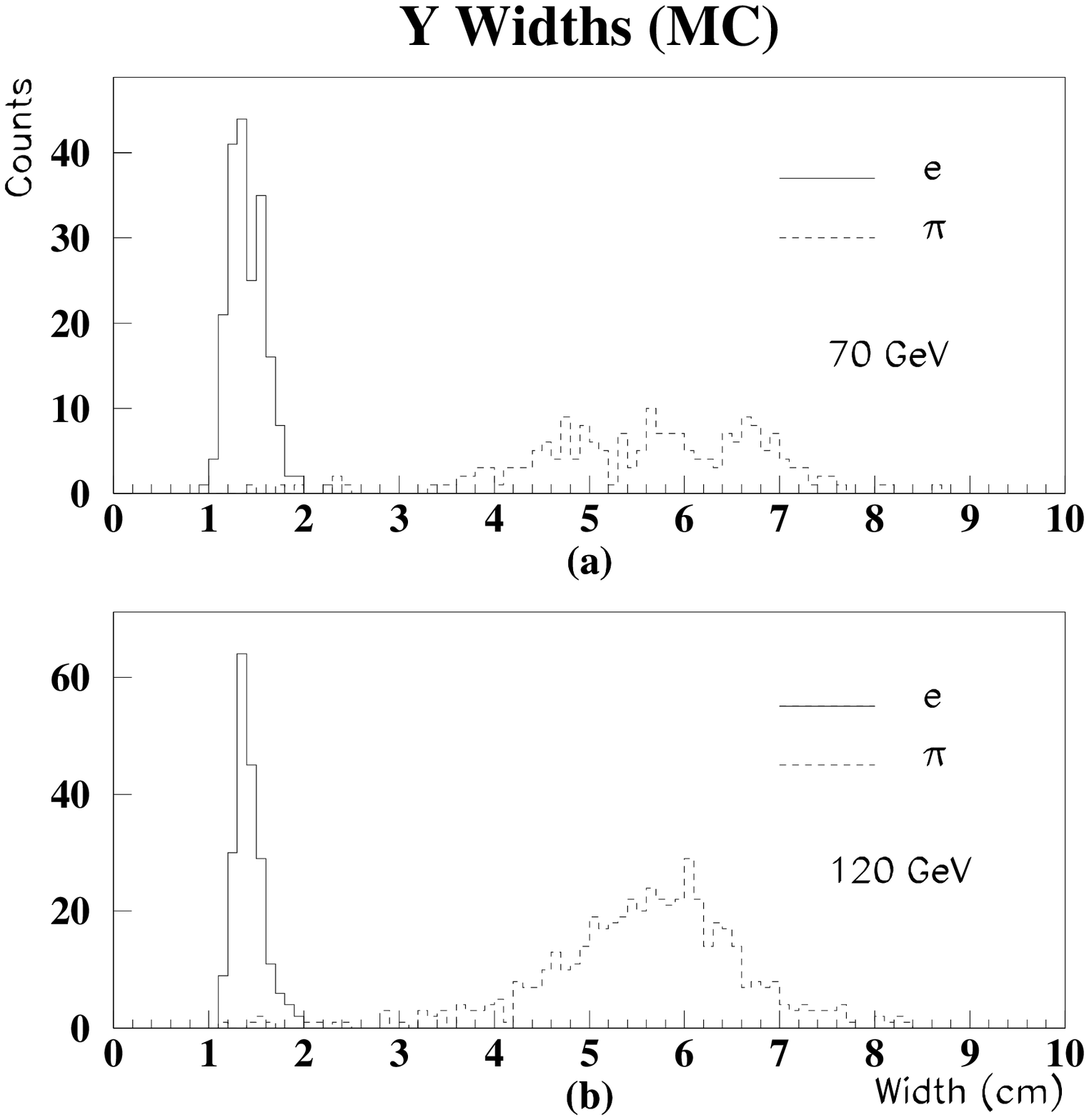}}
\caption{The histograms are Monte Carlo simulations of
the width distributions, with linear weights,
for (a) 70 and (b) 120 GeV electrons and pions
incident on tower 6.
See Fig.~\ref{ehsep} for the corresponding plot made
with test beam data.}
\label{mc_ywidth}
\end{figure}

The simulated energy response for pions
as a function of position, shown
in Fig.~\ref{mc_eofy}, is in agreement with the
data shown in Fig.~\ref{eofy}.
Fig~\ref{mc_ywidth} shows simulations of
the vertical shower widths for 70 and 120 GeV electrons and
pions. The Monte Carlo results are narrower than the
data shown in Fig.~\ref{ehsep}.

\section{Summary}

We have designed and constructed a lead scintillator sandwich
calorimeter for the ZEUS experiment at HERA. The calorimeter
is divided into 5 cm vertical towers read out on two sides with
wavelength shifting light guides coupled to photomultiplier
tubes. The calorimeter was tested in the H6 beam line at CERN 
with 120 GeV electrons, muons, pions and protons. 
Electrons can be cleanly separated from hadrons using
the energy weighted vertical width of a shower.
At 120 GeV the calorimeter is slightly 
over compensating, with an electron to hadron response ratio
of 0.96 and an
energy resolution 6\% at 120 GeV.
The horizontal position resolution, measured by
charge sharing between the two sides, is 20 cm/$\sqrt E$,
assuming a $1/\sqrt E$ dependence;
the vertical position resolution, measured by energy sharing
between the towers, is 10 cm/$\sqrt E$.
By using energetic neutrons produced by proton interactions
in a lucite target, the
overall energy scale was determined to 1.5\%.

Since the spring of 1995, the calorimeter has operated
successfully in the HERA tunnel 105.6 m
downstream of the ZEUS detector on the zero degree line.

\section*{Acknowledgments}
We thank T.~Tymieniecka for modeling the FNC during the
design stage, and using FLUKA
to study its energy resolution and $e/h$ response. 
E.~Borsato, F.~Czempik, C.~Fanin, R.~Fernholz,
T.~Kiang, K.~Loeffler,
H.~Stehfest, V.~Sturm, and K.~Westphal
provided us with much help constructing the calorimeter,
shipping it to CERN, and installing it in the HERA tunnel.
M. Roseman helped with the PMT tests and at CERN.
We thank
M.~Brki\'{c} for his assistance in setting up the computer
readout system for the CERN tests. B.~Racky kindly arranged
for us the transport of our $^{60}$Co source to and from CERN.
We also thank F. Dittus
for making available the adjustable table on which we mounted
the calorimeter, and K.~Elsener for his invaluable assistance with
the H6 beam. We are also grateful to CERN for making
the beam test possible.
J.~Prentice helped with cosmic ray tests
during the initial stages.  
The ZEUS collaboration has been continually enthusiastic and
supportive, in particular, R.~Klanner, G.~Wolf and U.~Koetz.
We also thank the HERA machine group who 
helped install the FNC, and 
who provided the beam line modifications which greatly
enhance the performance for physics of the FNC.
Finally, we especially thank
the DESY directorate for the continual interest they have 
shown in the project, and for the financial support they
provided. 

\newpage

\clearpage

\clearpage

\clearpage

\clearpage

\clearpage

\end{document}